\documentclass[reprint,prb,twocolumn,groupedaddress]{revtex4-2}
\usepackage[T1]{fontenc}
\usepackage{subcaption}
\usepackage{graphicx}
\usepackage[font=small,labelfont=bf]{caption}
\usepackage{comment}
\usepackage{float}

\usepackage{amsfonts, amsmath, amssymb}
\usepackage{xcolor}
\usepackage{tikz}
\usepackage{wrapfig}
\usetikzlibrary{arrows.meta,positioning}
\usepackage{titlesec}
\definecolor{link1}{RGB}{178,24,43}
\definecolor{link2}{RGB}{49,54,149}
\definecolor{link3}{RGB}{214,96,77}

\PassOptionsToPackage{colorlinks=true, linkcolor=link1, citecolor=link1, urlcolor=link2}{hyperref}
\usepackage{hyperref}
\hypersetup{pdfborder={0 0 0}}

\titlespacing{\subsection}{5pt}{5pt}{5pt}
\titlespacing{\section}{2pt}{6pt}{6pt}
\titlespacing{\subsubsection}{5pt}{6pt}{6pt}

\begin{document}

\title{\textbf{Learning of Statistical Field Theories}}

\author{Shreya Shukla}
\email{sshukla@lanl.gov}
\affiliation{Theoretical Division, Los Alamos National Laboratory, Los Alamos, New Mexico 87545, USA}

\author{Abhijith Jayakumar}
\email{abhijithj@lanl.gov}
\affiliation{Theoretical Division, Los Alamos National Laboratory, Los Alamos, New Mexico 87545, USA}

\author{Andrey Y.\ Lokhov}
\email{lokhov@lanl.gov}
\affiliation{Theoretical Division, Los Alamos National Laboratory, Los Alamos, New Mexico 87545, USA}

\begin{abstract}
    Recovering microscopic couplings directly from data provides a route to solving the inverse problem in statistical field theories, one that complements the traditional—often computationally intractable—forward approach of predicting observables from an action or Hamiltonian. Here, we propose an approach for the inverse problem that uniformly accommodates systems with discrete, continuous, and hybrid variables. We demonstrate accurate parameter recovery in several benchmark systems—including Wegner's Ising gauge theory, $\phi^4$ theory, Schwinger and Sine-Gordon models, and mixed spin–gauge systems, and show how iterating the procedure under coarse-graining reconstructs full non-perturbative renormalization-group flows. This gives direct access to phase boundaries, fixed points, and emergent interactions without relying on perturbation theory. We also address a realistic setting where full gauge configurations may be unavailable, and reformulate learning algorithms for multiple field theories so that they are recovered directly using observables such as correlations from scattering data or quantum simulators. We anticipate that our methodology will find widespread use in practical learning of field theories in strongly coupled regimes where analytical tools might fail.
\end{abstract}

\maketitle

Statistical field theories underpin our understanding of collective behavior, from magnetic ordering to confinement in gauge theories~\cite{Wilson1974,Kogut1979,Zinn-Justin2002,Peskin1995,Goldenfeld1992,Wegner1971}. They trade microscopic detail for an effective description whose critical behavior falls into a  set of universality classes with system–independent scaling laws~\cite{Negele1988,Binney1992}. Choosing the right effective model, one can understand the mechanisms behind critical phenomena ubiquitous in nature. The usual method to fit a model to observed data works by solving a forward inference problem: the candidate models are first postulated, relevant quantities are computed from them using analytical or numerical methods and then compared against experimental observations. However, computing observables given an action is often a provably hard problem \cite{sinclair1989approximate, jerrum1993polynomial, goldberg_jerrum_2015} and this difficulty manifests itself often as computational bottlenecks, especially in the strong coupling regime.

In contrast, in this work we argue that the inverse task of learning the Hamiltonian directly from data can be surprisingly tractable, using recent methods where infeasible \cite{Wainwright2008,Mezard2009} explicit partition $\mathcal{Z}=\int\!\mathcal{D}\phi\,e^{-S[\phi]}$ function calculations are avoided \cite{Vuffray2022,Klivans2017}. This inverse view is especially powerful in strongly coupled regimes, where perturbation theory methods fail and rich physics such as high‑temperature superconductivity, quantum spin liquids, or quark confinement emerges. For systems with continuous degrees of freedom, Score Matching enables parameter learning without evaluating the partition function ~\cite{Hyvarinen2005, pabbaraju2023provable}. For discrete variables, Pseudolikelihood and Interaction Screening objectives offer an analogous route~\cite{Besag1975,Vuffray2022,Lokhov2018}. Thus, the parameter learning stage of the inverse problem can be substantially easier than the forward simulation. However, practical obstacles to applications in field theories still exist: experiments and many simulations provide only partial statistics—structure factors, low‑order correlators, rather than full field configurations; and real systems mix discrete spins, continuous fields, fermions and gauge links, defying any single estimator.

\begin{figure*}[ht]
    \centering
    \subfloat[]{\includegraphics[width=0.4\linewidth]{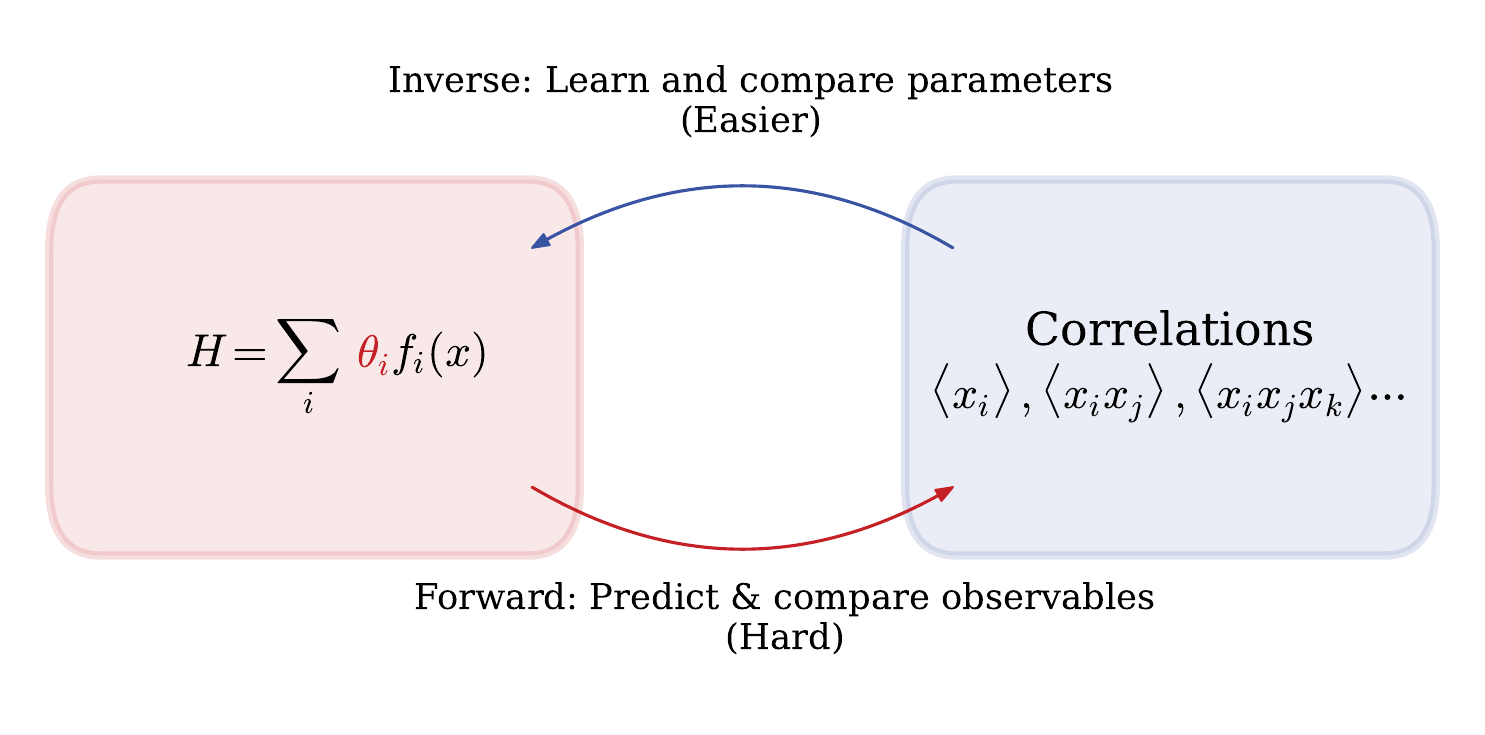}}
    \subfloat[1D Ising Model ($J = 2.0$ and $J = 0.2$) \label{fig:im_learned}]{
        \includegraphics[width=0.3\linewidth]{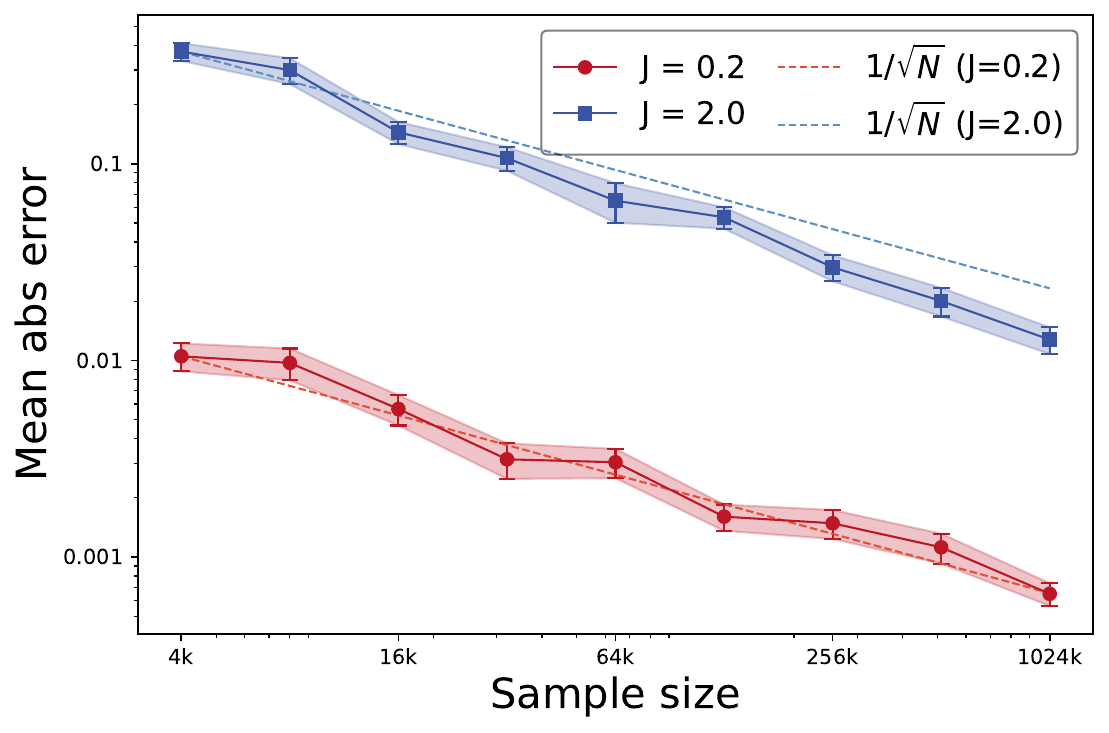}\label{fig:1dising_corr}
    }
    \subfloat[$\phi^4$ Theory ($\lambda = 2.0$ and $\lambda = 0.2$) \label{fig:phi4_learned}]{
        \includegraphics[width=0.3\linewidth]{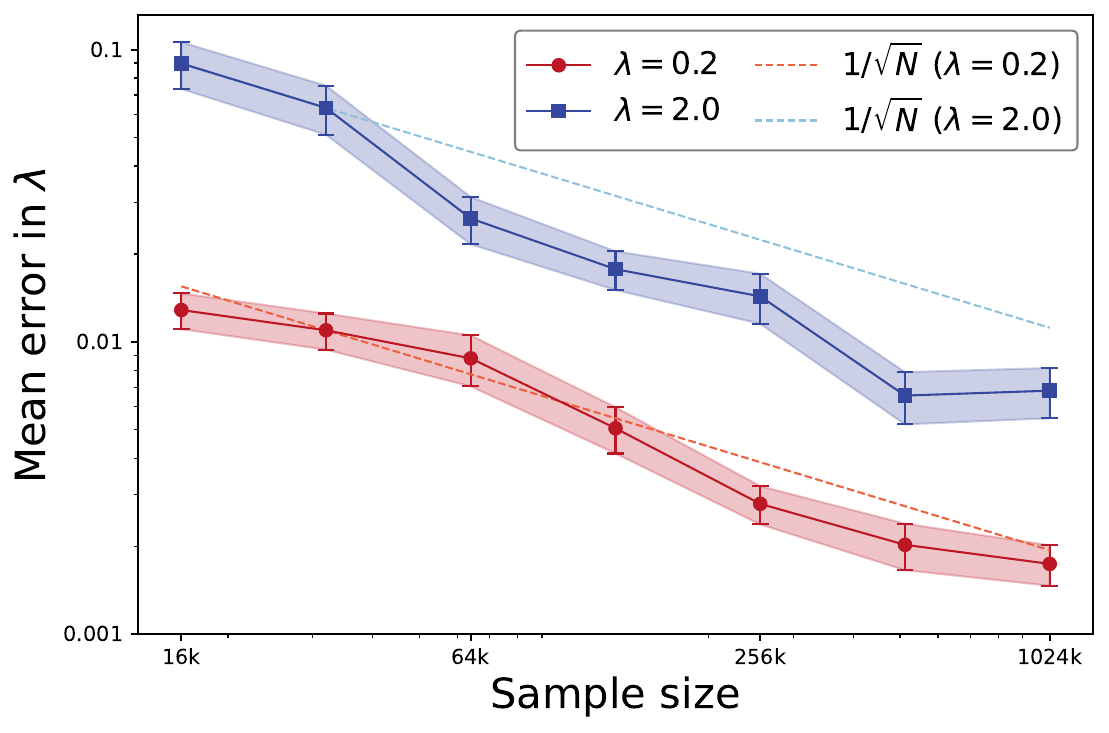}\label{fig:phi4_corr}
    }
    \caption{(a) \textbf{Schematic of forward and inverse approaches to model validation.} A traditional approach to validating a field theory would consist in computing an observable and comparing with an experiment, which is limited to tractable regimes of weak couplings. We put forward an alternative proposal, where model parameters are learned from data and compared to the postulated theory, an apporach which is not limited by the coupling strength. (b,c) \textbf{Learning from correlations.} Error plots for 1D Ising model and $\phi^4$ theory, learning from measured correlations for both large and small coupling values separated by order of magnitude and averaged over 20 repetitions at each step, demonstrating accurate and consistent parameter recovery for both discrete and continuous parameter models not only for weak coupling, but also in the case of strong coupling where perturbation theory fails.}
    \label{fig:ising_phi4_errors}
\end{figure*}

Beyond parameter recovery, one wants to know how theories evolve across scales, i.e.\ under renormalization-group (RG) transformations. A broad suite of numerical RG methods exists: Monte-Carlo RG~\cite{Swendsen1979,Swendsen1984}, functional RG with derivative truncations~\cite{Wetterich1993,Berges2002}, Density Matrix Renormalization Group
% DMRG/TNR and other
and other tensor-network approaches~\cite{White1992,Evenbly2015,Hauru2018}, hierarchical and multiscale wavelet methods~\cite{WCRG2023,lempereur2024hierarchic}, and recent ML-assisted techniques~\cite{Hou2023,Jho2023,Chung2021,Di_Carlo_2025}.
% Other ML approaches have also tackled this problem. Ref.~\cite{WCRG2023} developed a multiscale wavelet method for data-driven energy estimation and generation, while Ref.~\cite{lempereur2024hierarchic} introduced hierarchic flows for sampling high-dimensional probabilities.
Yet most of these track flows of observables or effective potentials rather than bare couplings; or linearize near criticality or adopt model-specific ansätze.
% ; or assume full configurations of a single degree-of-freedom type. 
% Ref.~\cite{Di_Carlo_2025}, for instance, shares our data$\rightarrow$coarse-grain$\rightarrow$relearn philosophy, but is restricted to the Ising model with discrete spins.

Our work instead provides a unified framework for learning couplings and recovering RG flows across diverse systems: discrete spins, continuous fields, gauge theories, and even hybrid models with mixed degrees of freedom. For the learning problem, we demonstrate that Score Matching, Pseudolikelihood, or the combination thereof can efficiently solve the inverse problem in a broad class of statistical field theories. We also investigate the issue that the original formulation of these methods requires the observation of full field configurations for learning the parameters in the model. We show that under practical scenarios that are relevant for physical systems, this learning task can be solved even when we only have access to a set of lower-order moments of the system. Indeed, for many models of interests such as discrete gauge models, $\phi^4$ theory, and the Schwinger model, we learn microscopic parameters directly from correlations by re-expressing Pseudolikelihood and Score Matching objectives exactly in terms of statistical moments, without introducing any approximations, and achieve accurate parameter recovery with errors below $1\%$. We also construct various real-space RG schemes to learn renormalization group flows in models ranging from simple Ising systems to complex gauge theories. Our non-perturbative reconstruction of Renormalization Group (RG) flows reveals phenomena such as the emergence of relevant effects at marginal points and precise mapping of phase boundaries.

The implications of this work extend beyond technical advances. For experimentalists, our methods provide a direct route from correlations measured at different energy scales to underlying field theories, bypassing the need for full microscopic characterization. For theorists, the ability to learn non-perturbative renormalization group flows opens new windows into our understanding of strongly-coupled theories. Our methods also provide a path to infer low-energy effective field theories given experimental data. In what follows, we develop methods for different types of random variables, and then apply them to extract RG flows in Ising, $\phi^4$, Schwinger, and Sine–Gordon, and mixed $\mathbb{Z}_2$–U(1) models. Throughout, we demonstrate not only efficient parameter recovery but also physical insights gained from data‑driven RG analysis.

\begin{figure*}[ht]
    \centering
    \subfloat[]{\includegraphics[width = 0.32\linewidth]{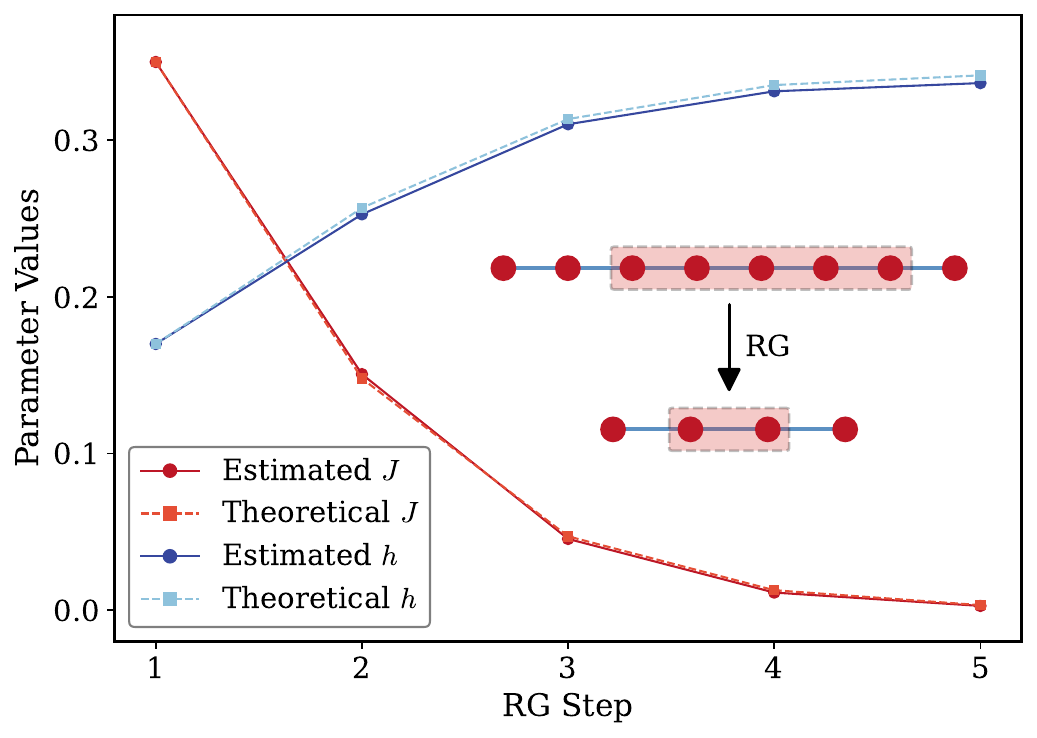}\label{fig:1dising_RG_all}}
    \subfloat[]{\includegraphics[width = 0.32\linewidth]{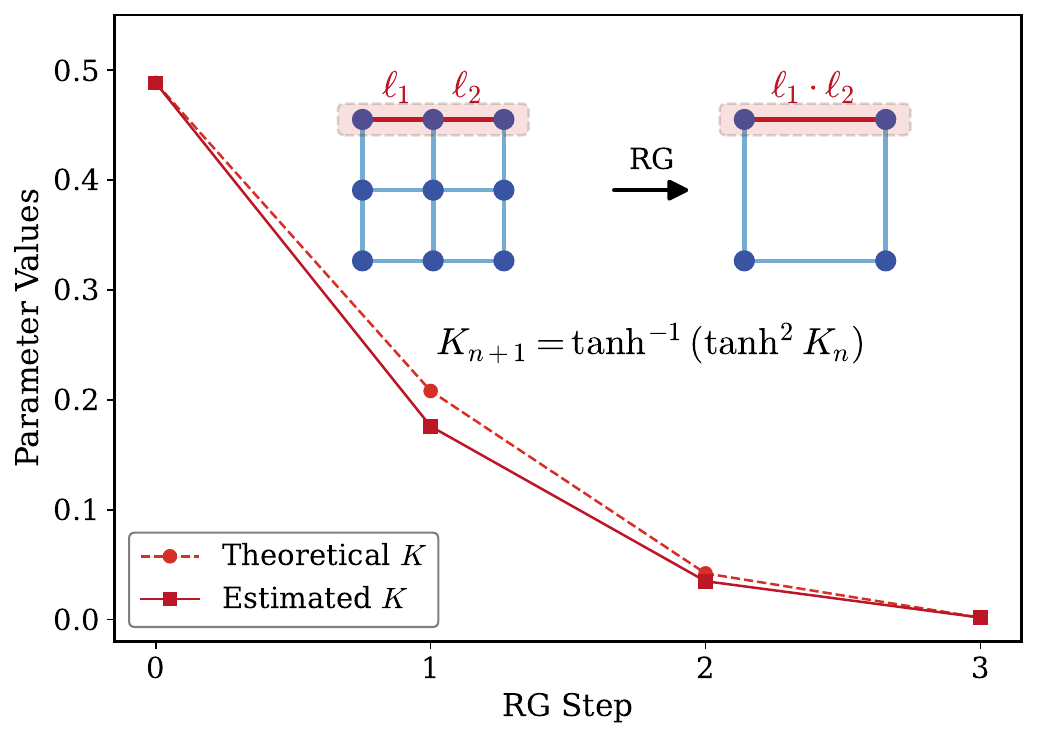}\label{fig:2dIGT_RG_all}}
    \subfloat[]{\includegraphics[width = 0.32\linewidth]{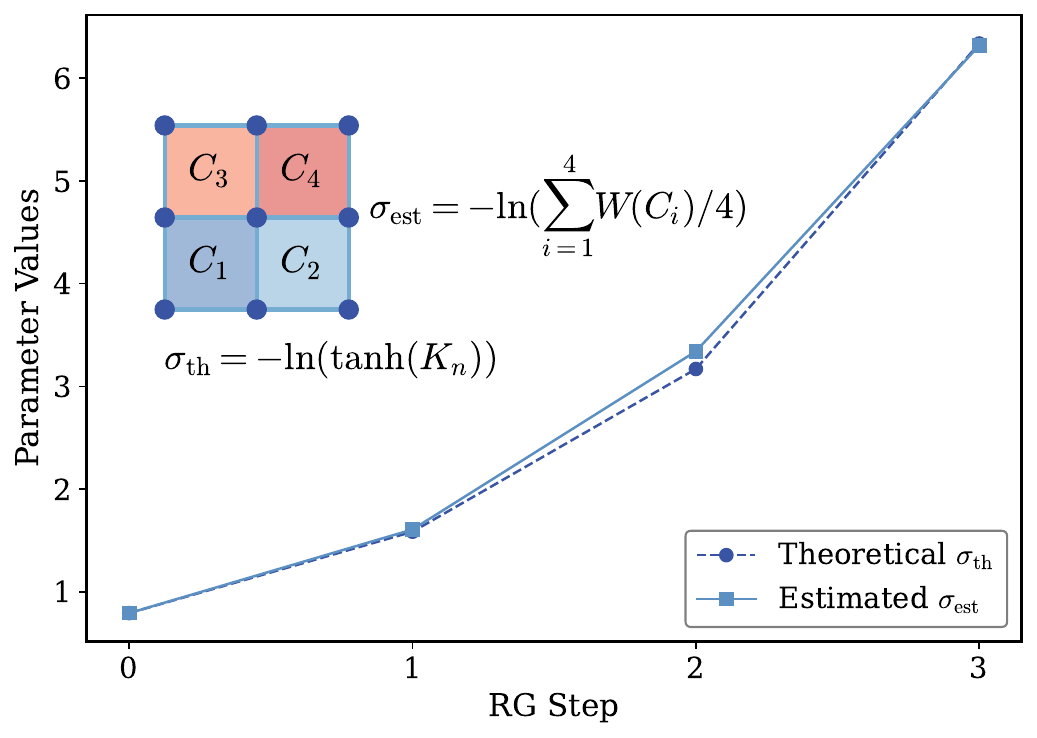}\label{fig:2dIGT_sigmaRG_all}}
    \caption{
        \textbf{Learning renormalization group (RG) flows in discrete lattice models.} 
        (a) For the 1D Ising model, we depict a schematic of the block-spin procedure, 
        and comparison between theoretical and learned RG trajectories for nearest-neighbor coupling $J$ and field $h$, 
        and
        (b) For the 2D Wegner's Ising gauge theory, we show a schematic of the coarse-graining procedure on the gauge links, the evolution of the theoretical and estimated normalized gauge coupling $K$ and (c) Schematic for estimating `String Tension' $\sigma$, with evolution of theoretical and estimated $\sigma$ under RG.
    }
    \label{fig:combined_RG}
\end{figure*}

\section{Discrete Models}
\label{sec:discrete}
\subsection{1D Ising Model}
We begin with the 1D Ising model to establish our methodology. The Hamiltonian takes the familiar form 
\begin{align*}
    H = -J \sum_{i=1}^L s_i s_{i+1} - h \sum_{i=1}^L s_i\;,
\end{align*} 
where $s_i \in \{-1, +1\}$ are spin variables, $J$ is the nearest-neighbor coupling, and $h$ is the external magnetic field, with periodic boundary conditions $s_{L+1} \equiv s_1$. This model describes magnetic systems ranging from one-dimensional quantum magnets realized in cold atom chains to effectively 1D magnetic materials~\cite{takayoshi2023phase}. The coupling $J$ is typically inferred from magnetic susceptibility measurements, while the field $h$ corresponds to applied magnetic fields measurable in magnetometry experiments.

For discrete variables, we employ pseudolikelihood estimation~\cite{Besag1975, Lokhov2018} to learn parameters from spin configurations. Detailed description of the method for generic models is provided in Appendix~\ref{sec:PL_dd}. This algorithm  works by maximizing the conditional likelihood of the model. Unlike the full likelihood, the computational tractability comes from the fact that conditionals are one-dimensional distributions which can be explicitly written down including their normalization. The resulting is still exact and consistent \cite{Lokhov2018}: in Fig.~\ref{fig:im_learned}, we demonstrate that this approach successfully recovers values for both weak ($J = 0.2$) and strong ( $J = 2.0$) couplings separated by an order of magnitude, with errors showing the expected $N^{-1/2}$ scaling as the amount of data increases. The robust performance across this range establishes pseudolikelihood as a reliable tool for discrete systems, setting the stage for our RG analysis.

\textbf{RG Analysis.} To extract RG flows, we implement Kadanoff's block-spin transformation~\cite{Kadanoff1966,Niemeijer1973}, grouping adjacent spins into block variables using majority voting: if both spins are equal, the block spin takes their value. If unequal, the block spin is $1$ or $-1$ at random, with probability $1/2$ for each.
% This transformation can be represented mathematically for block spins $\sigma_i$ as $P(\sigma_j | s_{2j-1}, s_{2j}) = \lim_{\alpha \to \infty} \exp[\alpha\sigma_j(s_{2j-1} + s_{2j})] / 2\cosh[\alpha(s_{2j-1} + s_{2j})]$, where the limit enforces the deterministic majority rule.
After marginalizing over original spins, the block-spin distribution retains the Ising form with renormalized couplings $(J', h')$ (see Appendix~\ref{sec:appendixA} for a detailed derivation). Fig.~\ref{fig:1dising_RG_all} shows that our learned RG trajectories match exact analytical results obtained from exact block spin (detailed in Appendix~\ref{sec:appendixA}) and analytical decimation in~\cite{Nauenberg_1975} with remarkable precision. The nearest-neighbor coupling $J$ decreases under coarse-graining as expected for this one-dimensional system, while the field $h$ flows according to its scaling dimension. This demonstrates that our method can discover which interactions matter at different scales without prior theoretical input.

\textbf{Learning 1D Ising Model from correlations.} A key insight enables learning even when only statistical moments are available rather than full configurations. For the homogeneous ($J_{ij} = J$ for all $i,j$) 1D Ising model, we can reformulate the pseudolikelihood objective entirely in terms of observable correlations. We define the magnetization $m = \langle s_i \rangle$, the nearest-neighbor correlation $c_1 = \langle s_i s_{i+1} \rangle$, and next-nearest correlation $c_2 = \langle s_i s_{i+2} \rangle$. By using careful algebraic manipulations (detailed in Appendix~\ref{sec:1dcorr_ising_appendix}), the loss function can be expressed entirely in terms of moments $m$, $c_1$ and $c_2$. This reformulation has profound implications: as demonstrated in Fig.~\ref{fig:1dising_corr}, we can learn model parameters directly from experimental measurements and without any approximations. In Appendix~\ref{sec:ps_general_d}, we derive a general formulation for $d-$dimensions, demonstrating the power of our approach. There we show that a $d-$dimensional homogeneous Ising model can be learned from only correlation functions up to order $2d$. Similar arguments can be easily extended to more general discrete models like Potts models. 
\\
\begin{figure*}[t]
  \centering
  \includegraphics[width = \linewidth]{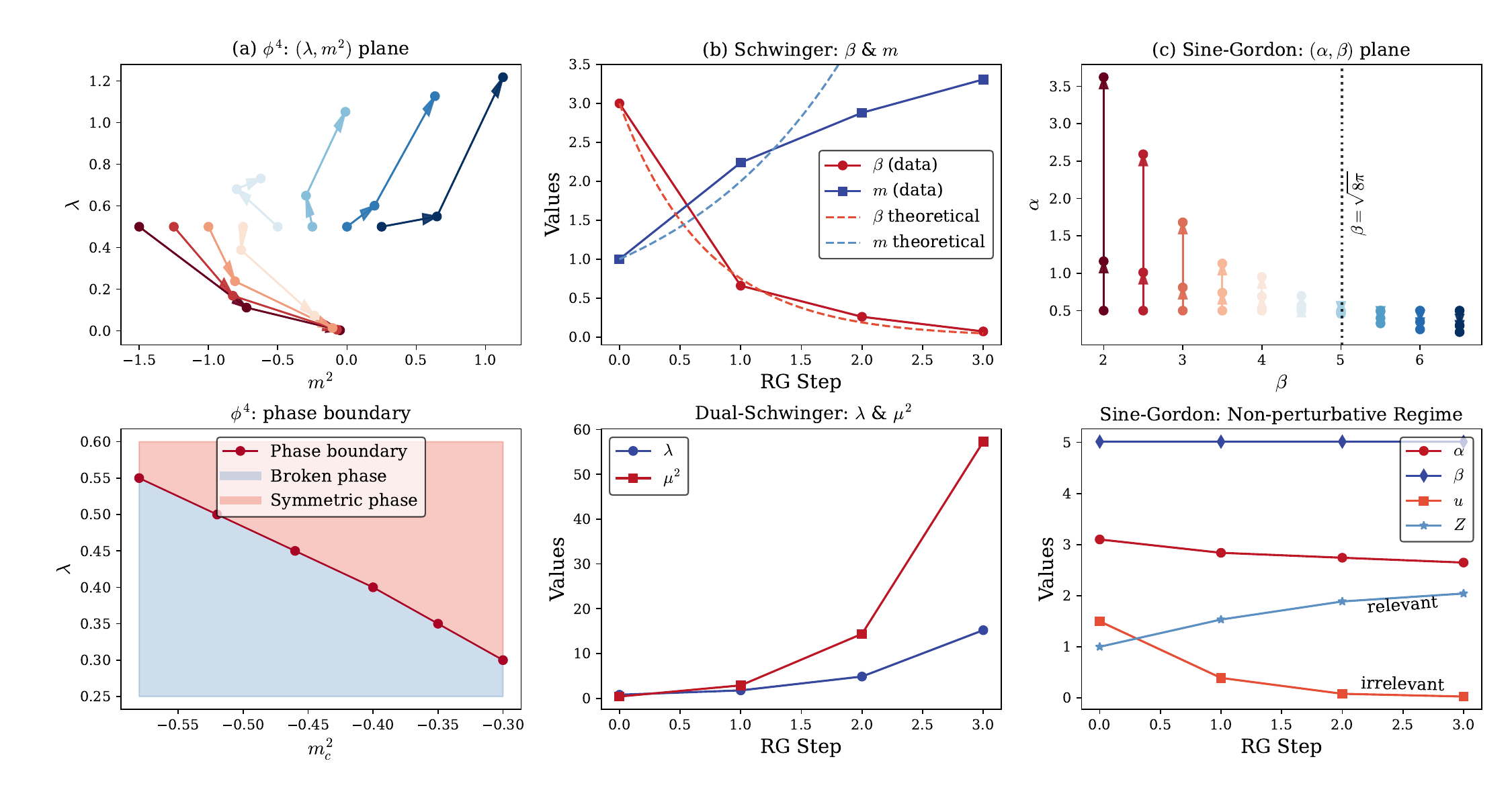}
  \caption{\textbf{Learning renormalization group (RG) flows in continuous models: $\phi^4$, Schwinger, and Sine–Gordon.}
  (a) Phase structure and representative RG flows in 2D $\phi^4$ theory.
  (b) Schwinger model and its dual: schematics with RG flows for $\beta,m$ with characteristic stalling when length scale $a \sim m_{\gamma}^{-1}$ and for $\lambda,\mu^2$.
  (c) Sine–Gordon: flows in the $(\alpha,\beta)$ plane and the evolution of $\alpha,\beta$, $u$(coefficient of higher harmonic term, decaying at low energies (IR)), $Z$ (wavefunction renormalization, growing in IR).}
  \label{fig:continuous_models_grid}
\end{figure*}

\subsection{2D Wegner's Ising Gauge Theory}
For this case, we work with the action:
\begin{align*}  
    H = -K \sum_{\square} \prod_{\ell \in \square} \sigma_{\ell}\;,  
\end{align*}  
where $\sigma_\ell \in \{\pm 1\}$ are binary variables or gauge fields living on lattice links, and the product runs around each elementary plaquette $\square$. This model, first introduced by Wegner~\cite{Wegner1971}, provides one of the simplest playgrounds for lattice gauge theory.  

A key observable in gauge theories is the expectation value of the Wilson loop $\langle W(C)\rangle$, where $W(C)$ is defined as the trace of the product of link variables around a closed contour $C$. Its expectation value encodes how charges interact: if it decays with area of the loop, the theory exhibits confinement: the potential barrier between two test charges increases linearly with distance. The proportionality constant in this exponential area law is called the string tension $\sigma$.

The 2D classical Wegner's Ising gauge theory is technically trivial: all plaquettes fluctuate independently and the system connects smoothly to the confined $K=0$ phase. However, it remains an ideal testbed for our methods because the behavior of Wilson loops and string tension is analytically tractable and directly illustrates confinement.

\textbf{RG Analysis.} Our real–space coarse graining maps the product of links on a larger plaquette to a single renormalized link in each direction for a smaller plaquette (Fig.~\ref{fig:2dIGT_RG_all}), yielding an effective coupling $K_{n}$ after every $n$th step.  The exact Migdal–Kadanoff prediction for a $2\times2$ block is~\cite{Kadanoff1976,Migdal1976} $K_{n+1} \;={\tanh}^{-1}(\tanh^2 K_n)$, and this theoretical recursion (dashed line in Fig.~\ref{fig:2dIGT_RG_all}) matches our learned flow, which drives $K$ toward the trivial fixed point (at zero) in the 2D $\mathbb{Z}_2$ gauge theory.  In Appendix~\ref{appendix:WIGT} demonstrate that the RG-scheme of multiplying coarse-links to form a fine link is gauge-covariant and local. We also introduce other real-space RG schemes that preserve gauge invariance, and show convergence of $K$ to the trivial fixed point.

Crucially, for the same initial coupling $K_0 = 0.45$, we extract the string tension $\sigma$ directly from Wilson--loop expectation values, independently of $K$. For a rectangular loop $C$ with area $A(C)$, we fit  $\langle W(C) \rangle = e^{-\sigma A(C)}$ to obtain $\sigma$ from the data. We then compare our results with the analytic prediction $\sigma(K_n) = -\ln\!\bigl(\tanh K_n\bigr)$~\cite{Wegner1971,Kogut1979}, finding excellent agreement(Fig.~\ref{fig:2dIGT_sigmaRG_all}). Since larger values of $\sigma$ correspond to stronger confinement, the growth of $\sigma$ under coarse graining in the 2D $\mathbb{Z}_2$ gauge theory provides a clear demonstration of how the strength of confinement increases along the RG flow.

\section{Continuous Models}
\label{sec:continuous}
\subsection{2D \texorpdfstring{$\phi^4$}{} Model}

The $\phi^4$ model serves as our first example of continuum field theory. This is also the simplest model used to demonstrate continuous phase transitions, and describes everything from the liquid-gas critical point to the Higgs mechanism in particle physics~\cite{Peskin1995}. On a 2D square lattice with unit spacing, the Euclidean Hamiltonian reads
\begin{align*}
H = \sum_{\mathbf{x}} \left[
    \frac{1}{2} \sum_{\mu=1}^2 (\phi_\mathbf{x} - \phi_{\mathbf{x}+\hat{\mu}})^2 
    + \frac{1}{2} m^2 \phi_\mathbf{x}^2 
    + \frac{\lambda}{4} \phi_\mathbf{x}^4 
\right],
\end{align*}
where $\mathbf{x} = (x,\tau) \in \mathbb{R}^2$, $\phi_{\mathbf{x}} \in \mathbb{R}$ is a scalar field,  $m^2$ is the mass parameter (which can be negative, signaling instability of the symmetric vacuum), and $\lambda > 0$ ensures stability of the potential at large field values.

We use the Advanced HMC package~\cite{xu2020advancedhmc} to generate samples. For continuous fields, we use score matching to learn parameters~\cite{Hyvarinen2005}. Score matching works by matching the score of the data (i.e the gradient of the energy function) to score of the model. For distributions over continuous variables in the Gibbs form, this is an effective way to fit models to statistical data while avoiding partition function computations. More details about score matching can be found in Appendix~\ref{sec:SM_cd}.

Our results demonstrate accurate parameter recovery across a wide range of couplings. As shown in Fig.~\ref{fig:phi4_learned}, we successfully learn $\lambda$ values spanning an order of magnitude.

\textbf{RG Analysis and Phase Boundary.} The $\phi^4$ theory exhibits a rich phase structure that our method captures directly from data. The system undergoes a transition between a symmetric phase where $\langle\phi\rangle = 0$ and a broken-symmetry phase where the field acquires a non-zero expectation value. Traditional approaches locate this transition through mean-field theory or perturbative RG, but our data-driven method maps the phase boundary without such approximations.

For coarse-graining, we take the median value of $\phi$ on each block and then rescale the resulting configuration so that its total kinetic energy matches that of the original system.

Fig.~\hyperref[fig:continuous_models_grid]{3a} reveals the phase structure in detail. These results are consistent with prior determinations of the critical line~\cite{LoinazWilley1998}. Deep in the broken phase where $m^2 \ll -\lambda$, both couplings flow toward zero while maintaining $m^2 < 0$, indicating the system flows to a trivial theory. In the symmetric phase, both $\lambda$ and $m^2$ increase under RG, reflecting the growing stability of the symmetric vacuum as we probe larger scales. Close to the phase boundary, the flow nearly stalls — $m^2$ barely shifts across RG steps, showing that the system is near a critical point.

\textbf{Learning \texorpdfstring{$\phi^4$}{} theory from correlations.} A key insight is that the score matching objective can be expressed entirely in terms of statistical moments of the field configurations. One may work with full samples, but in many experimental/simulation contexts, the only available data is correlations. In \ref{subsec:momentbasedphi4}, we show the moment-based score matching objective in terms of correlations involving upto sextic terms $\langle \phi^6\rangle$ and terms involving nearest neighbors. Fig.~\ref{fig:err_corr-phi4} shows that the errors for parameters $\lambda$ and $m^2$ scale as $N^{-1/2}$, confirming that score-based learning from correlations achieves the statistical efficiency of direct configuration-based methods without requiring access to the full field data.

However, we note that this approach works best when couplings enter linearly in the Hamiltonian. For nonlinear interactions (such as when
the parameters appear inside trigonometric functions), the objective cannot be written purely in
terms of low-order moments. In such cases, one must use full field configurations rather than correlation data alone.

\textbf{Sample Complexity Analysis.} From Figs.~\ref{fig:lambda_complexity} and \ref{fig:m_complexity} one can see that the learning complexity grows exponentially and polynomially respectively, with coupling strength, following $N_\lambda \sim \exp(0.65\lambda)$ and $N_{m^2} \sim 1758 (m^2)^3$ for fixed target accuracy.  While the exponential scaling with $\lambda$ reflects the increasing difficulty  of learning strongly-coupled systems, the complexity of the learning task even in these regimes scales only polynomially with the number of sites in the system. This is in sharp contrast to the complexity of computing observables from a given model (the forward direction in \figureautorefname~\ref{fig:ising_phi4_errors}), which scales exponentially with the number of sites in the strong coupling regime. %of sampling from strongly-coupled distributions, learning remains tractable.
\\
\subsection{2D Massive Schwinger Model (QED\texorpdfstring{$_2$}{}) and its Dual}
Consider the 2D-Schwinger Model with 2 flavors~\cite{Schwinger1962, shimizu2014grassmann} and Wilson Fermions:
\begin{align*}
    H &= -\sum_{\text{plaq}}\beta(1-\cos(\theta_{\text{plaq}})) + \sum_f\psi^{(f)\dagger}\bigg((m+2)\psi^{(f)}(x)\\
    &\qquad -\frac{1}{2}\sum_{\mu = 1}^{2} U_{\mu}(x)(1-\gamma_\mu)\psi^{(f)}(x + \hat{\mu}) \\
    &\qquad + U_{\mu}^{\dagger}(x-\hat{\mu})(1 + \gamma_\mu) \psi^{(f)}(x - \hat{\mu}) \bigg)\;,
\end{align*}
where $\theta_{\text{plaq}}$ is the plaquette composed of gauge fields $U_\mu \in \text{U}(1)$ on links, and $\psi$ are fermion fields on sites. We also work with the model dual to the 1 flavor case~\cite{Coleman1976}:
\begin{align*}
    H_{\text{D}} &= \sum_{\mathbf{x}}\left[\sum_{\hat{\mu} = 1}^2\frac{(\phi_{\mathbf{x}} - \phi_{\mathbf{x} + \hat{\mu}})^2}{2} + \frac{\mu^2 \phi_{\mathbf{x}}^2}{2} - \lambda\, \text{cos}(\beta \phi_\mathbf{x} - \theta)\right]\;,
\end{align*}
where $\mathbf{x} = (x,\tau) \in \mathbb{R}^2$, $\theta \in [-\pi,\pi]$ is an angle parameter, $\mu^2$ is the mass and $\phi_{\mathbf{x}} \in \mathbb{R}$ is a scalar field. The 2D Schwinger model or quantum electrodynamics in 1+1 dimensions, provides a valuable theoretical laboratory for understanding gauge theories with interesting phase diagrams, as studied in~\cite{Dempsey2024}. While not directly realizable in condensed matter, it captures essential features of confinement also present in QCD.

Using results from \hyperref[sec:sf_sm]{Appendix B}, we formulate the score function and learn the parameter values from samples. We implement the Schwinger model simulations for data generation using the \texttt{JulianSchwingerModel.jl} package~\cite{Lin2023}. The errors in the Schwinger Model decrease with sample size, but due to expensive computation time, their scaling was not verified. The errors in dual-Schwinger Model can be verified to scale as $N^{-1/2}$.
    
\textbf{RG Analysis.} At large length scales the massive Schwinger model confines: the effective gauge coupling $g$ increases, driving the lattice parameter $\beta \sim g^{-2}$ downward. Our learned RG flow (Fig.~\ref{fig:continuous_models_grid}b) shows that the fermion mass $m$ initially grows, as expected from dimensional scaling. At later steps, however, this growth slows relative to the additive mass renormalization  $m\propto a$ for Wilson fermions~\cite{Angelides2023,Dempsey2022} with the flattening setting in near the scale $a\sim m_\gamma^{-1}$. While this might partially be attributed to lattice effects, the fact that this slowdown is observed at $a \sim m_\gamma^{-1}$ suggests that once the photon becomes heavy, non-perturbative dynamics suppress further renormalization of $m$ beyond dimensional scaling(Fig.~\hyperref[fig:continuous_models_grid]{3b}).

For the two-flavor Schwinger model, the exact bosonized theory contains two scalar fields, while our single-scalar model is only an effective proxy. Consequently, the parameter identifications $\mu^2 \sim \beta^{-1}$ and $\lambda \sim m$ are only approximate, and the flows in Fig.~\hyperref[fig:continuous_models_grid]{3b} should be read qualitatively rather than as exact one-to-one matches. With that caveat, the flows reproduce the expected infrared behavior: $\mu^2$ increases as $\beta$ decreases to zero, while $\lambda$ increases due to an increase in $m$.
\\
\subsection{2D Sine-Gordon Model}
We work with the Euclidean Hamiltonian
\begin{align*}
    H &= \sum_{\mathbf{x}}
    \left[\sum_{\hat{\mu} = 1}^2Z\cdot\frac{(\phi_{\mathbf{x}} - \phi_{\mathbf{x} + \hat{\mu}})^2}{2}+ \alpha(1-\text{cos}(\beta\phi_{\mathbf{x}}))\right]\;,
\end{align*}
where $\mathbf{x} = (x,\tau) \in \mathbb{R}^2$, $\phi_{\mathbf{x}} \in \mathbb{R}$ is a scalar field, the lattice spacing is $1$ and $Z$ is the wavefunction renormalization term. The Sine-Gordon model is an important example of an integrable quantum field theory~\cite{Zamolodchikov1978,Faddeev1987} exhibiting rich phase structure, including the Kosterlitz-Thouless transition~\cite{Kosterlitz1973,Jose1977}.

\textbf{RG Analysis.} The cosine interaction in this model corresponds to the real part of the vertex operator $\mathcal{V}(\phi) = \mathrm{e}^{:i\beta\phi:}$, whose scaling dimension at the Gaussian fixed point is $\Delta = \beta^2 / 4\pi$. In two dimensions, this operator is relevant for $\Delta < 2$, irrelevant for $\Delta > 2$, and marginal at $\Delta = 2$~\cite{DJAmit_1980,oak2017exact}, corresponding to $\beta^2 = 8\pi$. While this classification is valid only near the fixed point, our numerical results show that $\Delta$ remains a qualitatively accurate predictor even for large values of $\alpha$: for $\beta^2 < 8\pi$, the coupling $\alpha$ grows under coarse-graining, driving the system toward a gapped phase, while for $\beta^2 > 8\pi$, it decreases and the theory flows to a gapless state (Fig.~\hyperref[fig:continuous_models_grid]{3c}). 

The marginal case $\beta_0^2=8\pi$ is particularly revealing. Perturbatively, both $\alpha$ and $\beta$ are stationary to first order, yet we observe $\alpha$ slowly decaying to zero along the RG trajectory (Fig.~\ref{fig:continuous_models_grid}f). We also added a second harmonic term to the Hamiltonian
of the form $u\,\cos(2\beta \phi_{x,t})$ and find it to be infrared–irrelevant in agreement with the analysis of Amit \emph{et al.}~\cite{DJAmit_1980} that higher harmonics are not generated at long distances.

\begin{figure*}[ht]
    \centering
    % First row: RG Flows
    \includegraphics[width=\linewidth]{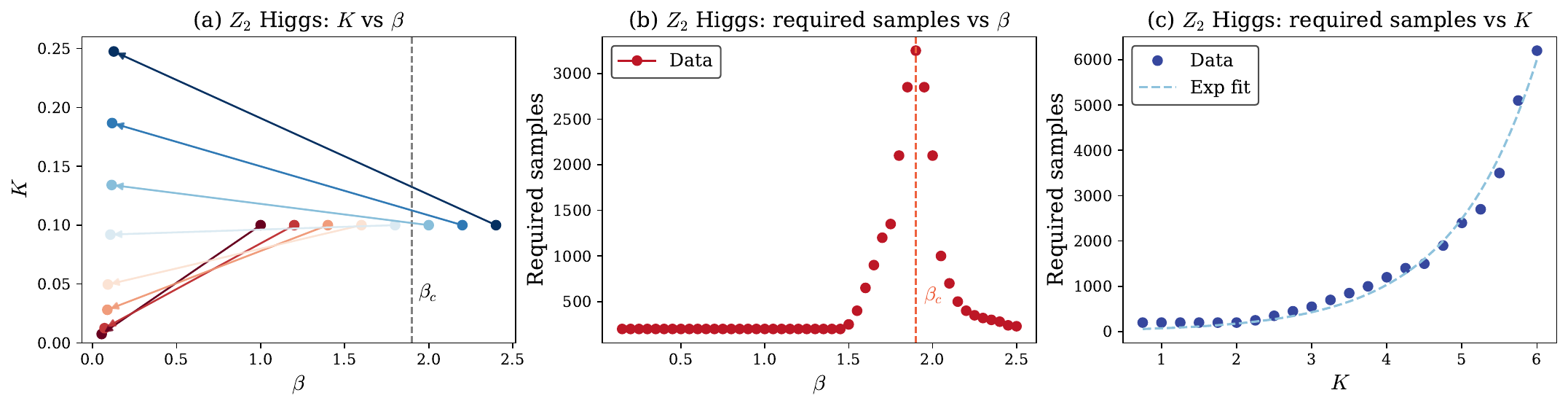}
    \caption{
        \textbf{RG Flows and Complexity Plots for Mixed Model parameters $\beta$ and $K$ for one step.} (a) RG Flow for $\beta$ and $K = 0.1$, with a non-trivial fixed point at  $\beta$ close to $1.8$. (b) Complexity for $\beta$ versus sample size at $K = 0.1$. (c) Complexity for $K$ with exponential fit versus sample size.}
    \label{fig:Z2_higgs}
\end{figure*}

Finally, the wavefunction renormalization term $Z$ proves indispensable. Forcing $Z=1$ along RG flow produces a non-monotonic running of $\alpha$ at the marginal point, indicating that the kinetic term must flow for internal consistency. Allowing $Z$ to run, we find that it increases from unity for all initial conditions, underscoring that wavefunction renormalization effects cannot be ignored in a faithful RG reconstruction.
\\
\section{Mixed Model}
\label{sec:mixed}
\subsection{A hybrid \texorpdfstring{$\mathbb{Z}_2$}{}--U(1) Theory}

Real physical systems often involve multiple types of degrees of freedom involving complex interactions. To demonstrate our framework's versatility, we study a model combining discrete Ising spins with continuous U(1) gauge fields. While not gauge-invariant, this model is inspired by analogous gauge-Higgs systems that have been extensively studied and exhibit rich phase diagrams~\cite{Fradkin1979}. 

The Hamiltonian couples spins on sites to gauge fields on links:
\begin{align*}
    H = -K \sum_{x,\mu,\nu} &U_{\mu}(x)U_{\nu}(x + e^{\mu})U^{\dagger}_{\mu}(x + e^{\nu})U^{\dagger}_{\nu}(x) 
    \\ \qquad &- \beta \sum_{x,\mu}\sigma(x)\text{Re}{(U_{\mu}(x))}\sigma(x + e^{\mu})\;,
\end{align*}
where $\sigma \in \{-1, +1\}$ are Ising variables,  $U_{\mu} \in \text{U}(1)$ are link variables, $e^\mu$ are unit vectors in direction $\mu$ and the first sum runs over elementary plaquettes. The plaquette term describes gauge dynamics, while the second term represents matter-gauge coupling.

The mixed alphabet necessitates a hybrid learning strategy that combines our discrete and continuous techniques. We first learn $\beta$ using pseudolikelihood on the spin degrees of freedom, treating the gauge fields as fixed external fields. The conditional probability $P(\sigma_x | \{\sigma_y, U\}) \propto \exp[\beta \sigma_x \sum_{y \sim x} \text{Re}[U] \sigma_y]$ is the only term with spins $\sigma$, allowing its isolation. With $\beta$ determined, we then apply score matching to learn the gauge coupling $K$ from the continuous link variables.

This sequential scheme recovers both parameters with the expected $N^{-1/2}$ error scaling. The success of this hybrid approach suggests a general strategy for complex systems: decompose the learning problem according to the structure of degrees of freedom, then combine specialized techniques for each component. The same strategy readily generalizes to gauge–Higgs systems, fermion–boson mixtures, and other composite theories across condensed‑matter and high‑energy contexts.

\textbf{RG Analysis.} To coarse-grain the data from this model, we use our standard procedure: we map the product of links to a link in every direction and use majority rule for spins on sites. The sample‑complexity curve in Fig.~\ref{fig:Z2_higgs}b exhibits a sharp maximum near $\beta_c \simeq 1.9$, signalling criticality.  Consistently, the one–step flow in the $(\beta,K)$ plane (Fig.~\ref{fig:Z2_higgs}a) shows that, in the vicinity of this point, $K$ switches from decreasing to increasing under coarse‑graining—i.e. the $\beta$‑function for $K$ changes sign, confirming the presence of a non‑trivial fixed point.

\textbf{Sample Complexity Analysis.} For fixed $K=0.1$, we determine the number of samples needed to reach an absolute error below $0.01$ as a function of $\beta$. The pronounced peak around $\beta\!\sim\!1.9$ (Fig.~\ref{fig:Z2_higgs}b) reflects the critical window: long-range correlations reduce the effective Monte-Carlo sample size (via large autocorrelation time $\tau_{\text{int}}$), maximizing the samples needed for a fixed error. However, we note that \emph{given sufficiently independent samples}, the learning part of the algorithm remains unaffected. The dependence on $K$ over our explored range on the other hand, is smooth; an exponential fit gives $N_K \sim \mathrm{e}^{0.88 K}$ for achieving the same error threshold (Fig.~\ref{fig:Z2_higgs}c).\\

\section*{Conclusions and Outlook}
In this work, we put forward the idea that learning of field theories provides an alternative to forward-simulation approaches that become challenging or intractable at strong coupling.  We have shown that it is possible to learn couplings and RG flows directly from data in a range of statistical field theories. Using pseudolikelihood for discrete systems, score matching for continuous ones, and a combination of estimators for hybrid models, we obtained accurate parameter estimates and reconstructed RG flows that agree with known analytical and numerical results, and capture non-perturbative behavior in the cases where analytical results are not known. The ability to formulate these estimators in terms of correlation functions in several relevant discrete and continuous theories makes the methods suitable even when only partial statistics are available, which is often the case with the available experimental data.

Looking ahead, there are several directions that are naturally open to exploration. An immediate path forward involves expanding our methods to fundamental and effective field theories of strong interactions, as well as testing estimators developed in this work directly on experimental data, for example neutron scattering measurements or quantum simulator outputs. In our numerical experiments, the choice of coarse-graining steps were model-specific rather than graph-agnostic; is it possible to design a data-driven approach that would discover the coarse-graining automatically or as a part of learning procedure? A longer-term goal would be to leverage recent theoretical progress in learning from non-equilibrium \cite{BreslerGamarnikShah2018, dutt2021exponential, GaitondeParity2025} and metastable \cite{jayakumar2025discretedistributionslearnablemetastable} states to extend the framework to non-equilibrium dynamics, which would open the door to studying driven systems and time-dependent effective theories.

\section*{Code Availability}
All simulation and learning code used in this work is publicly available at
\href{https://github.com/lanl-ansi/LearningFieldTheories}{github.com/lanl-ansi/LearningFieldTheories} \cite{2025LFT}.\\

\textbf{Acknowledgments}
The authors acknowledge support from the U.S. Department of Energy/Office of Science Advanced Scientific Computing Research Program and from the Laboratory Directed Research and Development program of Los Alamos National Laboratory under Project No. 20240245ER.

\bibliographystyle{apsrev4-2}
\bibliography{ref}
\appendix
\onecolumngrid
\vspace{5ex}
\section{Learning algorithms}
\label{sec:LA}
\subsection{Pseudolikelihood for Discrete Data}
\label{sec:PL_dd}
\noindent Introduced by Besag~\cite{Besag1975}, pseudolikelihood provides a computationally tractable alternative to maximum likelihood estimation for discrete statistical models where the partition function is intractable. The key insight is to replace the joint likelihood with a product of conditional likelihoods, each of which can be computed exactly without knowledge of the global normalization.

\subsubsection{Mathematical Formulation}
\noindent Consider a discrete statistical model with joint probability distribution
\begin{align}
P(\mathbf{x} ; \boldsymbol{\theta}) = \frac{1}{Z(\boldsymbol{\theta})} \exp\left(-H(\mathbf{x} ; \boldsymbol{\theta})\right),
\end{align}
where $\mathbf{x} = (x_1, x_2, \ldots, x_N)$ represents the configuration of discrete variables, $H(\mathbf{x} ; \boldsymbol{\theta})$ is the energy function parameterized by $\boldsymbol{\theta}$, and $Z(\boldsymbol{\theta}) = \sum_{\mathbf{x}} \exp(-H(\mathbf{x} ; \boldsymbol{\theta}))$ is the intractable partition function. The pseudolikelihood replaces the joint distribution with the product of conditional distributions:
\begin{align}
PL(\boldsymbol{\theta}) = \prod_{i=1}^{M} P(x_i | \mathbf{x}_{\setminus i}, \boldsymbol{\theta}),
\end{align}
where $\mathbf{x}_{\setminus i}$ denotes all variables except $x_i$. The crucial advantage is that each conditional probability can be computed exactly:
\begin{align}
P(x_i | \mathbf{x}_{\setminus i}, \boldsymbol{\theta}) = \frac{\exp(-H(\mathbf{x} ; \boldsymbol{\theta}))}{\sum_{x_i'} \exp(-H(\mathbf{x}_{i \to x_i'} ; \boldsymbol{\theta}))},
\end{align}
where $\mathbf{x}_{i \to x_i'}$ denotes the configuration $\mathbf{x}$ with $x_i$ replaced by $x_i'$. The denominator involves summation only over the possible values of $x_i$, making it computationally feasible.

\subsubsection{Pseudolikelihood Estimator}

\noindent The pseudolikelihood estimator maximizes the log-pseudolikelihood:
\begin{align}
\hat{\boldsymbol{\theta}}_{PL} = \arg\max_{\boldsymbol{\theta}} \sum_{i=j}^{N} \sum_{i=1}^{M} \log P(x_i^{(t)} | \mathbf{x}_{\setminus i}^{(t)}, \boldsymbol{\theta}),
\end{align}
where $\{\mathbf{x}^{(t)}\}_{j=1}^{N}$ represents $N$ independent samples from the true distribution. This optimization can be performed efficiently using gradient-based methods. As $N \to \infty$ (number of samples), $\hat{\boldsymbol{\theta}}_{PL} \to \boldsymbol{\theta}_{\text{true}}$. The error scaling is typically $O((N)^{-1/2})$, which we observe empirically in our results.

\subsubsection{Application to Ising Models}
\noindent For the Ising model with Hamiltonian $H = -J\sum_{\langle i,j \rangle} s_i s_j - h \sum_i s_i$, the conditional probability becomes:
\begin{align}
P(s_i | \mathbf{s}_{\setminus i}) = \frac{\exp\left(s_i\left(h + J\sum_{j \sim i} s_j\right)\right)}{2\cosh\left(h + J\sum_{j \sim i} s_j\right)},
\end{align}
where the sum runs over neighbors of site $i$. This leads to the negative log-pseudolikelihood objective used throughout our discrete model analysis.

\subsubsection{Moment-Based Pseudolikelihood Estimator}
We utilize a key property of this likelihood estimator: for several discrete models, the objective can be expressed in terms of statistical moments of data. In particular, for the $d$-dimensional classical Ising Model we show in Appendices~\ref{sec:1dcorr_ising_appendix} and \ref{sec:ps_general_d} that the pseudolikelihood estimator is expressible entirely in terms of moments up to order dependent on the dimension $d$ of the model. This moment-based formulation enables parameter learning for experimental data, where only moments and correlations are available, as opposed to full samples.\\

\subsection{Score Matching}
\label{sec:SM_cd}
\noindent Score matching, introduced by Hyvärinen~\cite{Hyvarinen2005}, provides an elegant method for parameter estimation in continuous probability models and bypasses the need for computation of the partition function. The method is based on matching the score function (gradient of log-density) between the model and data distributions.

\subsubsection{Mathematical Formulation}

\noindent Consider a continuous probability model with density:
\begin{align}
p(\mathbf{x} ; \boldsymbol{\theta}) = \frac{1}{Z(\boldsymbol{\theta})} \exp(-H(\mathbf{x} ; \boldsymbol{\theta})),
\end{align}
where $E(\mathbf{x} | \boldsymbol{\theta})$ is the energy function and $Z(\boldsymbol{\theta})$ is the partition function. The score function is defined as:
\begin{align}
\mathbf{s}(\mathbf{x} ; \boldsymbol{\theta}) = \nabla_{\mathbf{x}} \log p(\mathbf{x} ; \boldsymbol{\theta}) = -\nabla_{\mathbf{x}} H(\mathbf{x} ; \boldsymbol{\theta}),
\end{align}
where the partition function cancels out in the gradient, making the score function tractable even when $Z(\boldsymbol{\theta})$ is unknown.

\subsubsection{Score Matching Objective}

\noindent The score matching objective minimizes the expected squared difference between the model score and the data score:
\begin{align}
J(\boldsymbol{\theta}) = \frac{1}{2} \mathbb{E}_{p_{\text{data}}(\mathbf{x})} \left[ \|\mathbf{s}(\mathbf{x} ; \boldsymbol{\theta}) - \mathbf{s}_{\text{data}}(\mathbf{x})\|^2 \right],
\end{align}
where $\mathbf{s}_{\text{data}}(\mathbf{x}) = \nabla_{\mathbf{x}} \log p_{\text{data}}(\mathbf{x})$ is the score of the true data distribution. Since the data score is typically unknown, Hyvärinen showed that this objective can be rewritten using integration by parts as:
\begin{align}
J(\boldsymbol{\theta}) = \mathbb{E}_{p_{\text{data}}(\mathbf{x})} \left[ \sum_{i=1}^{d} \left( \frac{\partial s_i(\mathbf{x} | \boldsymbol{\theta})}{\partial x_i} + \frac{1}{2} s_i(\mathbf{x} | \boldsymbol{\theta})^2 \right) \right] + \text{const},
\end{align}
where $s_i(\mathbf{x} | \boldsymbol{\theta})$ is the $i$-th component of the score function, and the constant term is independent of $\boldsymbol{\theta}$.

\subsubsection{Practical Implementation}

\noindent Given a dataset $\{\mathbf{x}^{(t)}\}_{t=1}^{T}$ sampled from the true distribution, the empirical score matching objective becomes:
\begin{align}
\hat{J}(\boldsymbol{\theta}) = \frac{1}{T} \sum_{t=1}^{T} \sum_{i=1}^{d} \left[ \frac{\partial s_i(\mathbf{x}^{(t)} ; \boldsymbol{\theta})}{\partial x_i} + \frac{1}{2} s_i(\mathbf{x}^{(t)} ; \boldsymbol{\theta})^2 \right].
\end{align}

\noindent The score matching estimator is:
\begin{align}
\hat{\boldsymbol{\theta}}_{SM} = \arg\min_{\boldsymbol{\theta}} \hat{J}(\boldsymbol{\theta}).
\end{align}
This optimization can also be performed efficiently using gradient-based methods. As usual, when $N \to \infty$ (number of samples), $\hat{\boldsymbol{\theta}}_{SM} \to \boldsymbol{\theta}_{\text{true}}$. The error scaling we observe in this case is also $O((N)^{-1/2})$.
\subsubsection{Application to Field Theories}

\noindent For $d$-dimensional field theories with Hamiltonians of the form:
\begin{align}
H[\phi] = \sum_{\mathbf{x}} \left[ \frac{1}{2} \sum_{\mu = 1}^{d} (\phi_{\mathbf{x}} - \phi_{\mathbf{x}+\hat{\mu}})^2 + V(\phi_{\mathbf{x}}) \right],
\end{align}
where $V(\phi)$ is the potential, the score function at field point $\mathbf{x}$ is:
\begin{align}
s_{\mathbf{x}}[\phi] = -\frac{\delta H}{\delta \phi_{\mathbf{x}}} = \sum_{\mu = 1}^d (\phi_{\mathbf{x}+\hat{\mu}} + \phi_{\mathbf{x}-\hat{\mu}}) - 2d\phi_{\mathbf{x}} - V'(\phi_{\mathbf{x}}),
\end{align}
where $V'(\phi) = dV/d\phi$ is the derivative of the potential. Then the score matching objective becomes:
\begin{align}
J(\boldsymbol{\theta}) = \sum_{t=1}^{T} \sum_{\mathbf{x}} \left[ \frac{\partial s_{\mathbf{x}}[\phi^{(t)}]}{\partial \phi_{\mathbf{x}}^{(t)}} + \frac{1}{2} (s_{\mathbf{x}}[\phi^{(t)}])^2 \right],
\end{align}
which can be evaluated directly from field configurations without knowledge of the partition function.

\subsubsection{Moment-Based Score Matching}
\label{subsec:momentbasedphi4}
\noindent  A key insight utilized in our work is that for many field theories, the score matching objective can be expressed entirely in terms of statistical moments of the field configurations. This allows parameter learning even when only correlation functions are available rather than full field configurations. For example, in the $\phi^4$ theory, the score function involves terms like $\phi^3$, $\phi \Delta \phi$, and their derivatives. Through careful algebraic manipulations, these can be rewritten in terms of moments such as $\langle \phi^2 \rangle$, $\langle \phi^4 \rangle$, $\langle \phi^6 \rangle$, and nearest-neighbor correlations $\langle \phi_{\mathbf{x}} \phi_{\mathbf{x}+\hat{\mu}} \rangle$, given by
\begin{align}
    J_{\text{mom.}}(m^2, \lambda) = V \Big[ &\frac{1}{2} \big( (4 + m^2)^2 \langle \phi^2 \rangle + \lambda^2 \langle \phi^6 \rangle
+ 4 \langle \phi_{\text{n.n.}}^2 \rangle + 12 \langle \phi_{\text{n.n.}_1} \phi_{\text{n.n.}_2} \rangle\notag\\
&- 8(4 + m^2) \langle \phi \phi_{\text{n.n.}} \rangle  - 8\lambda \langle \phi^3 \phi_{\text{n.n.}} \rangle
+ 2\lambda(4 + m^2) \langle \phi^4 \rangle \big)\notag\\ 
&- (m^2 + 4) - 3\lambda \langle \phi^2 \rangle \Big]\;,
\end{align}
where $\phi_x$ is the field at site $x$, $\phi_{\text{n.n.}}\equiv\phi_{x+\hat\mu}$ is the field at a nearest neighbor of $x$, $\text{n.n.}_1,\text{n.n.}_2$ are two distinct nearest neighbors of the same $x$, $\langle\cdot\rangle$ denotes ensemble averages and $V=N^2$ is the number of sites. This moment-based formulation enables parameter inference from experimental data where only specific correlation functions are measurable, significantly broadening the practical applicability of the method. 
\begin{figure*}[t]
    \centering
    \subfloat[]{\includegraphics[width=0.4\linewidth]{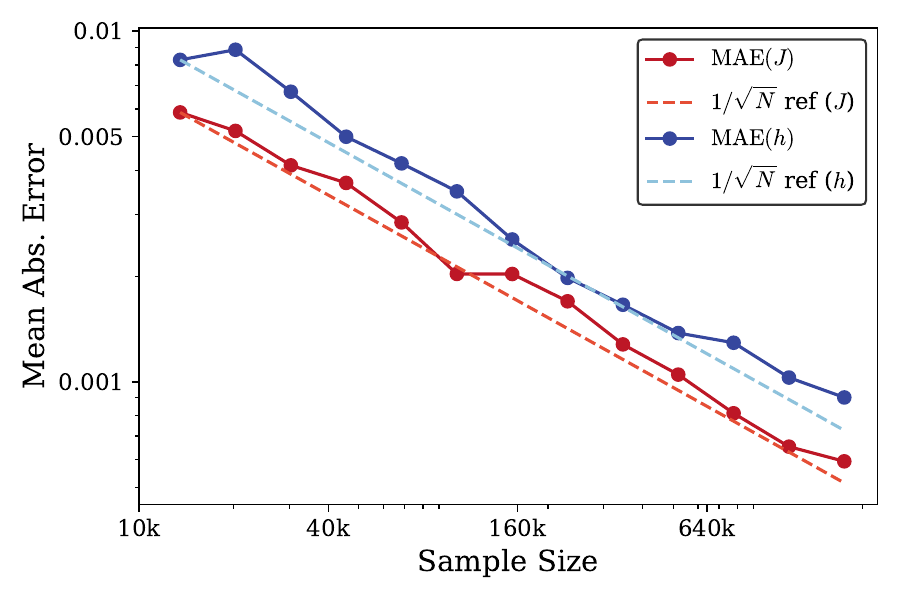}}\hspace{2ex}
    \subfloat[]{\includegraphics[width=0.4\linewidth]{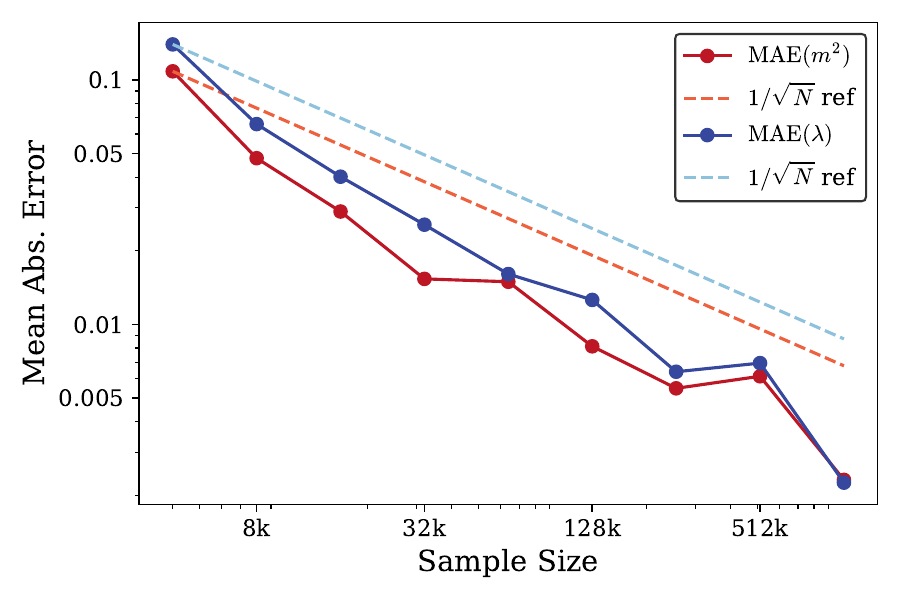}\label{fig:err_corr-phi4}}
    \caption{\textbf{Learning from correlations only:}
    (a) {1D Ising}: learning the couplings $J$ and $h$ from moments alone,
    (b) {2D $\phi^4$}: learning the parameters $\lambda$ and $m^2$ from correlations only. In both cases, the errors scale as $1/\sqrt{N}$ with the number of samples $N$, demonstrating efficient learning.}
\end{figure*}
\section{Methods}
\label{sec:methods}
\subsection{Sampling Methods}
\noindent We work with square space--time lattices equipped with periodic boundary conditions. For continuous models, spatial and temporal derivatives are discretized as finite differences on the lattice, so that kinetic terms take the standard nearest--neighbor form. We then employ different Monte Carlo algorithms for sampling optimized for each model class:
\begin{itemize}
\item \textbf{Discrete and mixed models} (Ising, Gauge, Mixed $\mathbb{Z}_2$--U(1)): Metropolis algorithm with single--spin or single--link updates, supplemented by cluster updates where efficient.
\item \textbf{Continuous field theories} ($\phi^4$, Sine--Gordon): Hamiltonian Monte Carlo (HMC) using the \texttt{AdvancedHMC.jl} package~\cite{xu2020advancedhmc}.
\item \textbf{Gauge theories with fermions} (Schwinger model): Hybrid Monte Carlo with pseudofermion fields using \texttt{JulianSchwingerModel.jl}~\cite{Lin2023}.
\end{itemize}
\noindent For each model, we generate $10^5$--$10^6$ thermalized configurations, with sufficient Monte Carlo steps between saved configurations to reduce autocorrelations.

\subsection{Real-Space Coarse-Graining Schemes}
\noindent We implement model-specific coarse-graining procedures that preserve the essential physical content while reducing degrees of freedom:\\

\paragraph{Ising Spin Models:}
For Ising spin systems, we implement Kadanoff's block-spin transformation~\cite{Kadanoff1966}. Spins on $2 \times 2$ blocks are mapped to single block spins using the majority rule:
\begin{align}
\sigma'_X = \begin{cases}
+1 & \text{if } \sum_{i \in \text{block}(X)} \sigma_i > 0 \\
-1 & \text{if } \sum_{i \in \text{block}(X)} \sigma_i < 0 \\
\pm 1 \text{ (random)} & \text{if } \sum_{i \in \text{block}(X)} \sigma_i = 0
\end{cases}
\end{align}
When the block sum is zero, the block spin is assigned randomly with equal probability. This transformation preserves the Ising form of the Hamiltonian while generating effective couplings between block spins.\\

\paragraph{Scalar Field Theories ($\phi^4$, Sine-Gordon, Dual Schwinger):} 
We apply a block-spin transformation that groups field values on $2 \times 2$ spatial blocks. For each block, the coarse-grained field value is taken as the median of the four constituent field values:
\begin{align}
\phi'_X = \text{median}\{\phi_{2X}, \phi_{2X+\hat{x}}, \phi_{2X+\hat{y}}, \phi_{2X+\hat{x}+\hat{y}}\},
\end{align}
where $X$ labels coarse-lattice sites. To maintain physical consistency, we rescale the coarse-grained configuration to preserve the total kinetic energy:
\begin{align}
\langle (\nabla \phi')^2 \rangle_{\text{coarse}} = \langle (\nabla \phi)^2 \rangle_{\text{fine}}.
\end{align}
This kinetic energy equalization ensures that the gradient terms in the action remain properly normalized across RG steps.\\

\paragraph{Pure Gauge Theories (Ising Gauge):}
For gauge link variables $\sigma_\ell \in \{\pm 1\}$, we implement the gauge-covariant procedure detailed in Appendix~\ref{appendix:WIGT}. Each coarse link is obtained as the product of fine links along the corresponding direction:
\begin{align}
\sigma'_{X,\mu} = \sigma_{2X,\mu} \cdot \sigma_{2X+\hat{\mu},\mu},
\end{align}
preserving local gauge invariance and Wilson loop observables.\\

\paragraph{Gauge Theories with Matter (Schwinger Model):}
The Schwinger model requires separate treatment for gauge and fermion degrees of freedom:
\begin{itemize}
\item \textbf{Gauge links}: Product of gauge links as above, $U'_{X,\mu} = U_{2X,\mu} \cdot U_{2X+\hat{\mu},\mu}$
\item \textbf{Pseudofermion fields}: Since we work with scalar pseudofermions rather than spinors for computational efficiency, we apply the same (complex) median + kinetic energy rescaling procedure used for scalar fields.
\end{itemize}

\subsection{Renormalization Group Transformations}

\noindent Our RG analysis follows a unified philosophy: generate equilibrium samples using appropriate Monte Carlo methods, learn microscopic parameters, apply real-space coarse-graining transformations, and re-learn effective parameters from the renormalized configurations. This iterative procedure reconstructs RG flows without relying on perturbative approximations.

The complete RG procedure for each model follows these steps:
\begin{enumerate}
\itemsep=2pt
\item \textbf{Initialize}: Start with bare parameters $\boldsymbol{\theta}_0$ and lattice size $L_0$ and lattice spacing $a$
\item \textbf{Sample}: Generate equilibrium configurations $\{\phi^{(i)}\}$ using appropriate Monte Carlo
\item \textbf{Learn}: Apply pseudolikelihood (discrete) or score matching (continuous) to extract $\boldsymbol{\theta}_0$
\item \textbf{Coarse-grain}: Apply model-specific transformation to obtain $\{\phi'^{(i)}\}$ on lattice $L_0/2$ with spacing $2a$
\item \textbf{Re-learn}: Extract effective parameters $\boldsymbol{\theta}_1$ from coarse-grained data
\item \textbf{Iterate}: Repeat steps 4-5 to trace the full RG trajectory $\boldsymbol{\theta}_0 \to \boldsymbol{\theta}_1 \to \boldsymbol{\theta}_2 \to \cdots$
\end{enumerate}

This procedure maps out complete RG flows in parameter space, revealing fixed points, phase boundaries, and the emergence of relevant operators across energy scales. This method does not depend on couplings being in the perturbative regime; As such, it captures strong-coupling physics that traditional analytical approaches cannot access.

\subsection{Phase Boundary Detection}

\noindent To map phase boundaries in the $\phi^4$ theory, we systematically scan the parameter space $(m^2, \lambda)$ and identify critical transitions through RG flow analysis. The phase boundary is located by detecting "turnaround points" where the RG flow direction changes—specifically, where the sign of the mass parameter flow $\Delta m^2 = m^2_{\text{RG}} - m^2_{\text{initial}}$ switches under coarse-graining. Our algorithm proceeds as follows:
\begin{enumerate}
\itemsep = 2pt
\item \textbf{Grid scan}: For each $\lambda$ value, we scan over a range of $m^2$ values with fine spacing $\delta m^2$
\item \textbf{RG trajectory}: At each point $(m^2_i, \lambda)$, we:
    \begin{enumerate}
    \itemsep = 1pt
    \item Generate $N$ equilibrium configurations using Hamiltonian Monte Carlo
    \item Apply $n_{\text{steps}}$ successive RG transformations
    \item Learn effective parameters at each step to trace the full RG trajectory
    \item Record the final mass parameter $m^2_{\text{final}}$ after RG evolution
    \end{enumerate}
\item \textbf{Flow direction analysis}: Compute the flow direction $\Delta m^2_i = m^2_{\text{final}} - m^2_i$ for each initial point
\item \textbf{Transition detection}: Identify adjacent points $(m^2_i, m^2_{i+1})$ where 
\begin{align}
\text{sign}(\Delta m^2_i) \neq \text{sign}(\Delta m^2_{i+1})
\end{align}
\item \textbf{Critical point estimation}: Locate the critical mass as $m^2_c \approx (m^2_i + m^2_{i+1})/2$
\end{enumerate}
\noindent This method exploits the fact that the symmetric and broken-symmetry phases exhibit opposite RG flow behaviors: in the symmetric phase, $m^2$ typically increases under coarse-graining (stabilizing the symmetric vacuum), while in the broken phase, $m^2$ becomes more negative (deepening the symmetry-breaking potential). The phase boundary corresponds precisely to the parameter values where this flow behavior transitions.

\subsection{Sample Complexity Analysis}

\noindent To characterize the statistical efficiency of our parameter learning algorithms, we perform systematic sample complexity studies. For each model and parameter regime, we determine the minimum number of samples $N_{\text{min}}$ required to achieve a target precision in parameter estimation. The procedure involves:
\begin{enumerate}
\itemsep=2pt
\item Fix target parameters $\boldsymbol{\theta}_{\text{true}}$ and precision threshold $\sigma_{\text{target}}$
\item For each parameter value of interest:
    \begin{enumerate}
    \itemsep=1pt
    \item Initialize sample size $N = N_{\text{start}}$
    \item Generate $R$ independent datasets of size $N$ from the true model
    \item Apply the learning algorithm to each dataset to obtain estimates $\hat{\boldsymbol{\theta}}^{(i)}$
    \item Compute the standard error: $\sigma_{\text{est}} = \text{std}(\hat{\boldsymbol{\theta}}^{(1)}, \ldots, \hat{\boldsymbol{\theta}}^{(R)}) / \sqrt{R}$
    \item If $\sigma_{\text{est}} \leq \sigma_{\text{target}}$, record $N_{\text{min}} = N$; otherwise, increase $N \rightarrow 1.1 \times N$ and repeat
    \end{enumerate}
\item Plot $N_{\text{min}}$ versus parameter values to reveal scaling relationships
\end{enumerate}

\noindent This analysis reveals how sample complexity scales with coupling strength and proximity to critical points, providing practical guidance for the number of configurations needed to achieve reliable parameter estimates in different physical regimes across all model classes studied.
\begin{figure}
    \subfloat[]{\includegraphics[width = 0.35\linewidth]{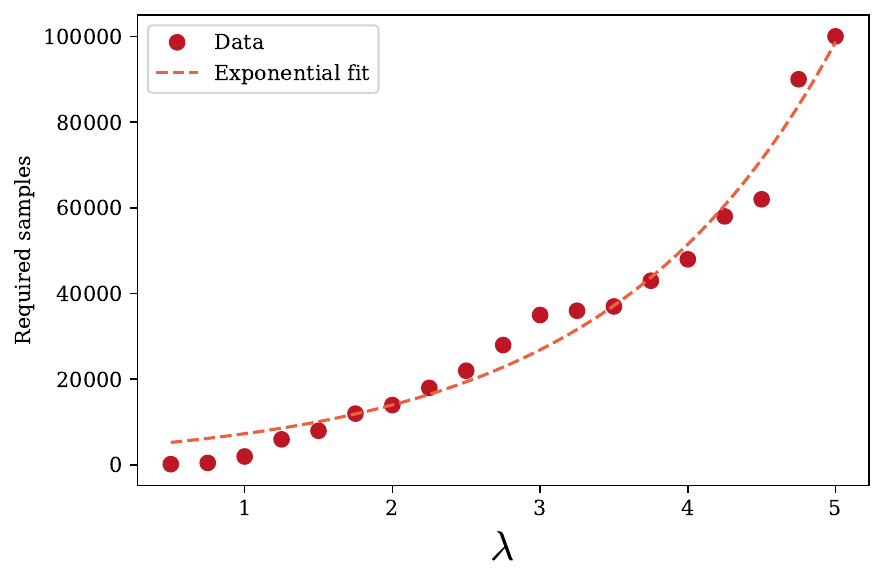}\label{fig:lambda_complexity}}
    \hspace{3ex}
    \subfloat[]{\includegraphics[width=0.35\linewidth]{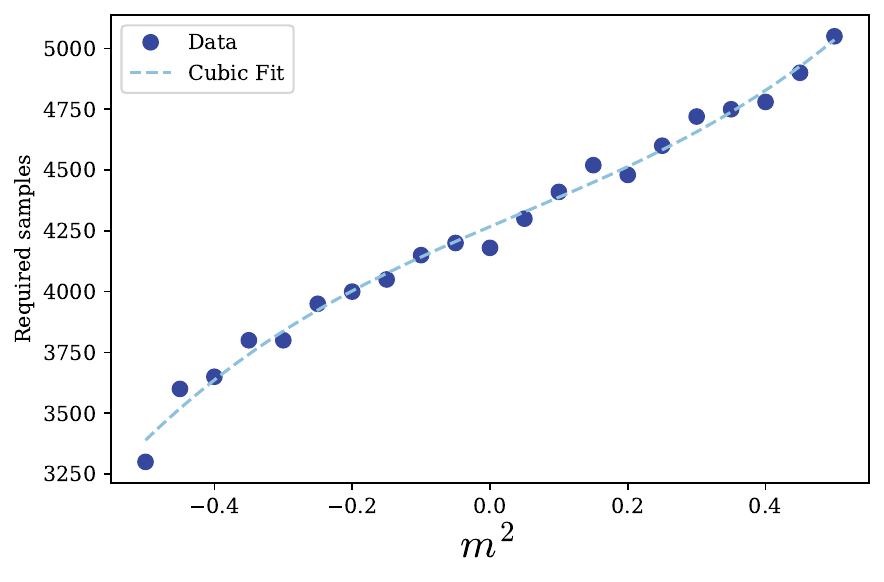}\label{fig:m_complexity}}
    \caption{(a-b) \textbf{Sample complexity for the $\phi^4$ Model:} Required number of samples as a function of $\lambda$ and $m^2$ to achieve a target error, with an exponential and cubic fit respectively.}
\end{figure}
\section{Block–Spin Probability for the 1D Ising Model}\label{sec:appendixA}
\noindent Consider a 1D Ising chain of even length $L$ with periodic boundary conditions $s_{L+1}\equiv s_1$.  We define block spins $\sigma_i$ from pairs $(s_{2i-1},s_{2i})$ using the probabilistic rule
\begin{align*}
p(\sigma_i=1\mid s_{2i-1},s_{2i}) &=
\begin{cases}
1   & s_{2i-1}+s_{2i}= +2,\\[2pt]
\frac{1}{2} & s_{2i-1}+s_{2i}= 0,\\[2pt]
0   & \text{otherwise},
\end{cases}\\[6pt]
p(\sigma_i=-1\mid s_{2i-1},s_{2i}) &=
\begin{cases}
1   & s_{2i-1}+s_{2i}= -2,\\[2pt]
\frac{1}{2} & s_{2i-1}+s_{2i}= 0,\\[2pt]
0   & \text{otherwise}.
\end{cases}
\end{align*}
Equivalently,
\begin{align}
    p(\sigma_i\mid s_{2i-1},s_{2i})
    =\lim_{\alpha\to\infty}
    \frac{e^{\alpha \sigma_i (s_{2i-1}+s_{2i})}}
    {2\cosh\!\big(\alpha (s_{2i-1}+s_{2i})\big)}\;.
    \label{eq:blockrule}
\end{align}
The marginal distribution of the block–spin configuration $\underline{\sigma}=\{\sigma_i\}$ is
\begin{align*}
p(\underline{\sigma})
= \sum_{\{s\}} p(\underline{\sigma}\mid \{s\})\,p(\{s\})
= \frac{1}{Z(J,h)} \sum_{\{s\}}
\prod_{j=1}^{L/2} p(\sigma_j\mid s_{2j-1},s_{2j})
\,e^{J\sum_{i=1}^{L} s_i s_{i+1} + h \sum_{i=1}^{L} s_i}.
\end{align*}
Using \eqref{eq:blockrule} and pulling the limit outside,
\begin{align*}
p(\underline{\sigma})
= \frac{1}{Z(J,h)} \lim_{\alpha\to\infty}
\sum_{\{s\}}
\exp\Bigg[
J\sum_{i=1}^{L} s_i s_{i+1}
+ h \sum_{i=1}^{L} s_i
+ \sum_{j=1}^{L/2}\Big(
\alpha \sigma_j (s_{2j-1}+s_{2j})
- \ln 2
- \ln\cosh\!\big(\alpha(s_{2j-1}+s_{2j})\big)
\Big)
\Bigg].
\end{align*}
Because $s_{2i-1},s_{2i}\in\{\pm1\}$, one can show that
\begin{align}
    \ln\cosh\!\big(\alpha (s_{2i-1}+s_{2i})\big)
    =\frac{1+s_{2i-1}s_{2i}}{2}\,\ln\cosh(2\alpha).
    \label{eq:cosh_id}
\end{align}
Substituting \eqref{eq:cosh_id} and reorganizing terms,
\begin{align*}
p(\underline{\sigma}) &=
\lim_{\alpha\to\infty}
\frac{(4\cosh 2\alpha)^{-L/4}}{Z(J,h)}
\sum_{\{s\}}
\exp\Bigg[
J\sum_{i=1}^{L} s_i s_{i+1}
+ h \sum_{i=1}^{L} s_i
+ \sum_{i=1}^{L/2}\Big(
\alpha \sigma_i (s_{2i-1}+s_{2i})
- \tfrac{1}{2} s_{2i-1}s_{2i}\ln\cosh(2\alpha)
\Big)
\Bigg].
\end{align*}
Splitting the interaction and field terms into odd–even and even–odd pairs,
\begin{align*}
J\sum_{i=1}^{L} s_i s_{i+1} + h \sum_{i=1}^{N} s_i =
J\sum_{i=1}^{L/2} s_{2i-1}s_{2i}
+ J\sum_{i=1}^{L/2} s_{2i}s_{2i+1}
+ h\sum_{i=1}^{L/2} (s_{2i-1}+s_{2i}).
\end{align*}
Hence
\begin{align*}
p(\underline{\sigma}) &=
\lim_{\alpha\to\infty}
\frac{(4\cosh 2\alpha)^{-L/4}}{Z(J,h)}
\sum_{\{s\}}
\exp\Bigg[
\left(J-\tfrac{1}{2}\ln\cosh(2\alpha)\right)
\sum_{i=1}^{L/2} s_{2i-1}s_{2i}
+ J\sum_{i=1}^{L/2} s_{2i}s_{2i+1}
+ \sum_{i=1}^{L/2} (h+\alpha \sigma_i)(s_{2i-1}+s_{2i})
\Bigg].
\end{align*}

\subsection*{Transfer Matrices}
\noindent Define two $2\times2$ transfer matrices acting on $(s=\pm1)$ states:
\begin{align*}
T_{2i-1,2i} &=
\lim_{\alpha\to\infty}\frac{1}{2}
\begin{pmatrix}
e^{J+2h+2\alpha\sigma_i}\,(\cosh 2\alpha)^{-1} & e^{-J}\\[2pt]
e^{-J} & e^{J-2h-2\alpha\sigma_i}\,(\cosh 2\alpha)^{-1}
\end{pmatrix},\\[8pt]
T_{2i,2i+1} &=
\begin{pmatrix}
e^J & e^{-J}\\[2pt]
e^{-J} & e^J
\end{pmatrix}.
\end{align*}
In the $\alpha\to\infty$ limit,
\begin{align*}
T_{2i-1,2i}(\sigma_i = +1) &=
\frac{1}{2}
\begin{pmatrix}
2e^{J+2h} & e^{-J}\\[2pt]
e^{-J} & 0
\end{pmatrix},\qquad
T_{2i-1,2i}(\sigma_i = -1) =
\frac{1}{2}
\begin{pmatrix}
0 & e^{-J}\\[2pt]
e^{-J} & 2e^{J-2h}
\end{pmatrix}.
\end{align*}
Therefore,
\begin{align}
p(\underline{\sigma})
= \frac{1}{Z(J,h)}\,
\mathrm{Tr}\big(T_{1,2}T_{2,3}\cdots T_{L-1,L}T_{L,1}\big).
\label{eq:trace_prod}
\end{align}
This trace is trivial to evaluate numerically for any configuration $\underline{\sigma}$. Assuming the block–spin distribution has the same Ising form,
\begin{align*}
p(\underline{\sigma})
=\frac{1}{Z(J',h')}
\exp\!\left[J'\sum_{i=1}^{L/2}\sigma_i\sigma_{i+1} + h'\sum_{i=1}^{L/2}\sigma_i\right],
\end{align*}
we can extract $(J',h')$ by evaluating \eqref{eq:trace_prod} for three configurations:
\begin{enumerate}
\item All $\sigma_i=+1$: call the trace $t_1$.
\item All $\sigma_i=-1$: call it $t_2$.
\item $\sigma_1=-1$, $\sigma_{2\ldots n}=+1$ (with $l=L/2$): call it $t_3$.
\end{enumerate}
Then
\begin{align*}
\frac{e^{J'l + h'l}}{Z(J',h')} &= t_1,\qquad
\frac{e^{J'l - h'l}}{Z(J',h')} = t_2,\qquad
\frac{e^{J'(l-4) + h'(l-2)}}{Z(J',h')} = t_3,
\end{align*}
and solving,
\begin{align}
J' &= \frac{1}{4}\ln\!\left(\frac{t_1}{t_3}\right) - \frac{1}{4l}\ln\!\left(\frac{t_1}{t_2}\right),\notag \\[4pt]
h' &= \frac{1}{2l}\ln\!\left(\frac{t_1}{t_2}\right).
\end{align}
\\
\section{Learning the 1D Ising Model from Two–Site Correlations}
\label{sec:1dcorr_ising_appendix}
\noindent The exact conditional distribution is
\begin{align*}
P(\sigma_i\mid \sigma_{i-1},\sigma_{i+1})
= \frac{\exp\left[\sigma_i\big(h + J(\sigma_{i-1}+\sigma_{i+1})\big)\right]}
{2\cosh\!\left(h + J(\sigma_{i-1}+\sigma_{i+1})\right)}.
\end{align*}
The per–site negative log pseudo–likelihood is
\begin{align*}
\mathcal{L}(J,h)
= -\frac{1}{L}\sum_{i=1}^{L}\log P(\sigma_i\mid\sigma_{i-1},\sigma_{i+1})
= -\frac{1}{L}\sum_{i=1}^{L}
\Big[\sigma_i\big(h + J(\sigma_{i-1}+\sigma_{i+1})\big)
- \log\big(2\cosh(h + J(\sigma_{i-1}+\sigma_{i+1}))\big)\Big].
\end{align*}
Introduce empirical moments
\begin{align}
m \equiv \frac{1}{L}\sum_i \sigma_i,\qquad
c_1 \equiv \frac{1}L\sum_i \sigma_i\sigma_{i+1},\qquad
c_2 \equiv \frac{1}{L}\sum_i \sigma_i\sigma_{i+2}.
\label{eq:define_moments}
\end{align}
The first term simplifies to $-h\,m - 2J\,c_1$. Let $S_i \equiv \sigma_{i-1}+\sigma_{i+1}\in\{+2,0,-2\}$ and denote
\begin{align*}
P_{+2}=P(S_i=+2),\quad
P_{0}=P(S_i=0),\quad
P_{-2}=P(S_i=-2),
\quad P_{+2}+P_{0}+P_{-2}=1.
\end{align*}
Match the first two moments:
\begin{align*}
E[S_i] &= 2(P_{+2}-P_{-2}) = \langle \sigma_{i} + \sigma_{i+1}\rangle = 2m,\\
E[S_i^2] &= 4(P_{+2}+P_{-2}) = \langle \sigma_{i}^2 + \sigma_{i+1}^2 + 2\sigma_{i}\sigma_{i+1}
\rangle = 2 + 2c_2,
\end{align*}
giving
\begin{align}
P_{+2}=\frac{1+2m+c_2}{4},\qquad
P_{0}=\frac{1-c_2}{2},\qquad
P_{-2}=\frac{1-2m+c_2}{4}.
\label{eq:moments_P}
\end{align}
Thus the second (log–cosh) contribution becomes
\begin{align*}
\frac{1+2m+c_2}{4}\,\log\!\big[2\cosh(h+2J)\big]
+ \frac{1-c_2}{2}\,\log\!\big[2\cosh(h)\big]
+ \frac{1-2m+c_2}{4}\,\log\!\big[2\cosh(h-2J)\big].
\end{align*}
\noindent\textbf{Final expression in terms of moments:}
\begin{align}
\mathcal{L}(J,h)
= - h\,m - 2J\,c_1
+ \frac{1+2m+c_2}{4}\log\!\big[2\cosh(h+2J)\big]
+ \frac{1-c_2}{2}\log\!\big[2\cosh(h)\big]
+ \frac{1-2m+c_2}{4}\log\!\big[2\cosh(h-2J)\big].
\end{align}

\section{Learning the \texorpdfstring{$d$}{}--dimensional Ising Model from \texorpdfstring{$2d$}{}--Site Correlations}
\label{sec:ps_general_d}
\noindent Consider the Ising model on a $d$–dimensional cubic lattice with $L^d$ spins $\{\sigma_i\}$, $\sigma_i\in\{\pm1\}$. If couplings are \emph{heterogeneous} (site/edge dependent), the conditional distinguishes each of the
$2^d$ neighbor \emph{configurations} $\sigma_{\partial i}\in\{\pm1\}^{2d}$, and the moment-only negative log pseudo–likelihood can be written as a weighted sum over these $2^d$ patterns (i.e., $2^d$ relations per site) rather than over the $2d{+}1$ values of $S_i$.
This gives a straightforward, dimension-agnostic extension of our correlation-based pseudolikelihood. In what follows, we will focus on the homogeneous case ($J_{ij}\equiv J$, $h_i\equiv h$) and its weights $P_s$ where we have uniform coupling $J$, and field $h$.  Each site $i$ has $2d$ nearest neighbors $j\sim i$. Define the local field
\begin{align*}
S_i \;\equiv\;\sum_{j\sim i}\sigma_j\;\in\;\{-2d,\,-2d+2,\,\dots,\,2d\}.
\end{align*}
The per–site negative log pseudo–likelihood is
\begin{align*}
\mathcal{L}(J,h)
=-\frac{1}{L^d}\sum_{i=1}^{L^d}
\log P\bigl(\sigma_i\mid\{\sigma_j\}\bigr)
=-\frac{1}{L^d}\sum_i
\Bigl[\sigma_i\,(h+J\,S_i)\;-\;\log\bigl(2\cosh(h+J\,S_i)\bigr)\Bigr].
\end{align*}
We use the identities
\begin{align*}
\mathbb{P}(\sigma_j=+1)=\frac{1+\sigma_j}{2},
\quad
\mathbb{P}(\sigma_j=-1)=\frac{1-\sigma_j}{2}.
\end{align*}
Suppose that $k$ of the $2d$-neighbors have spin $+1$, and the total spin $S_i = s$. Then 
\begin{align*}
S_i = s = \sum_{j=1}^{2d}\sigma_j
= \underbrace{(+1)+\cdots+(+1)}_{k\text{ times}}
  \;+\;\underbrace{(-1)+\cdots+(-1)}_{(2d-k)\text{ times}}
= k - (2d-k) = 2k - 2d.
\end{align*}
\begin{align*}
s = 2k - 2d
\quad\Longrightarrow\quad
k = \frac{s + 2d}{2}.
\end{align*}
Since \(s\) runs over the even values \(-2d,-2d+2,\dots,2d\), this \(k\) is always an integer. Then, if we define $A$ to be the indexing set of $+1$ spins of size $k$, $A\subseteq\{1,\dots,2d\}$ and the probability $P_s \equiv \mathbb{P}\bigl(\sum_j\sigma_j=s\bigr)$, we will have
\begin{align*}
P_s
&=\sum_{\substack{A\subseteq\{1,\dots,2d\}\\|A|=k}}
\mathbb{E}\!\Bigl[
\prod_{j\in A}\frac{1+\sigma_j}{2}
\;\prod_{j\notin A}\frac{1-\sigma_j}{2}
\Bigr]\\
&= 2^{-2d}\sum_{|A| = k}\mathbb{E}\Bigl[\prod_{j=1}^{2d}(1 + \epsilon_j \sigma_j)\Bigl]
\end{align*}
where $\epsilon_j = +1$ for $j \in A$ and $-1$ otherwise. Expanding the product over $j$, we get
\begin{align*}
    \prod_{i=1}^{2d}(1 + \epsilon_j \sigma_j) = \sum_{B \subseteq \{1,2\dots 2d\}}\left(\prod_{j \in B}\epsilon_j\right)\prod_{j \in B}\sigma_j
\end{align*}
where we define a new indexing set $B$. Inserting this expression, and swapping the sums 
\begin{align*}
P_s &= 2^{-2d}
       \sum_{B \subseteq \{1,\dots,2d\}}\,\Bigl\langle\prod_{j\in B}\sigma_j\Bigr\rangle
       \sum_{\substack{A\subseteq\{1,\dots,2d\}\\|A|=k}}\left(\prod_{j \in B}\epsilon_j\right)\\
       &= 2^{-2d}
       \sum_{B \subseteq \{1,\dots,2d\}} N_s(B)\,\Bigl\langle\prod_{j\in B}\sigma_j\Bigr\rangle\,.
\end{align*}
where we define $N_s(B)$ as
\begin{align*}
    N_s(B) := \sum_{\substack{A\subseteq\{1,\dots,2d\}\\|A|= k = \frac{(s+2d)}{2}}}\left(\prod_{j \in B}\epsilon_j\right) = \sum_{|A|=\frac{(s+2d)}{2}} (-1)^{|B\setminus A|}
\end{align*}
and counts the number of minus signs coming from indices in $B$ but not in $A$ (indices $\in A$ contribute $+1$.) Assuming translation invariance and defining,
\begin{align*}
m=\frac{1}{L^d}\sum_i\sigma_i,
\quad
c_1=\frac{1}{dL^d}\sum_{\langle ij\rangle}\sigma_i\sigma_j.
\end{align*}
where $c_1 = \langle \sigma_i \sigma_j \rangle$ is the average nearest-neighbor correlation. The negative log pseudo-likelihood loss can be expressed as
\begin{align*}
\mathcal{L}(J,h) = -h\,m - 2d\,J\,c_1 + \sum_{s = -2d}^{2d} 2^{-2d}
       \sum_{B \subseteq \{1,\dots,2d\}} N_s(B)\,\Bigl\langle\prod_{j\in B}\sigma_j\Bigr\rangle \log\big[2\cosh(h + Js)\big],
\end{align*}
This expression is {\em exact} for any lattice dimension $d$, involving correlators up to order~$2d$. Each $P_s$ is thus an explicit weighted sum of all spin correlation functions of order up to $2d$, with numerical coefficients that depend only on $d$ and $s$. This formulation allows the loss to be written entirely in terms of empirical moments, without approximation, provided all needed correlations are available from data. As an example, we verify this for the $d=1$ case. In this case, $S_i = \sigma_{1} + \sigma_{2} \in \{-2,0,+2\}$, where $\sigma_{i-1} \equiv \sigma_1$ and $\sigma_{i+1} \equiv \sigma_2$. For $S_i = +2$, $k=2 = 2d$, and hence the set $A$ has to be the full indexing set $\{1,2\}$. Thus, we have
\begin{align*}
    N_{+2}(B) = +1 \text{    for any set   }B \subseteq \{1,2\},\qquad\text{as}\quad B\setminus A = \emptyset\;.
\end{align*}
Substituting this into the relation for $P_s$ with $s = S_i = +2$,
\begin{align*}
    P_{+2} &= 2^{-2}\sum_{B \subseteq \{1,2\}}(+1)\cdot\langle \prod_{j\in B} \sigma_j \rangle \\
    &= \frac{1}{4}\left(1 + \langle\sigma_1\rangle + \langle\sigma_2\rangle + \langle\sigma_1\sigma_2\rangle\right) \\
    &= \frac{1}{4}\left(1 + 2m + c_2\right)\;,
\end{align*}
which matches our earlier results from~\ref{eq:moments_P}, and we use the notation for $m$ and $c_2$ from Eq.\ref{eq:define_moments} in the previous section. Although this notation might appear bulky for small values of $d$, it nonetheless provides a formal expression for larger $d$ values. 

\section{Wegner's Ising Gauge Theory}
\label{appendix:WIGT}

\begin{figure}[H]
    \centering
    {\includegraphics[width=\linewidth]{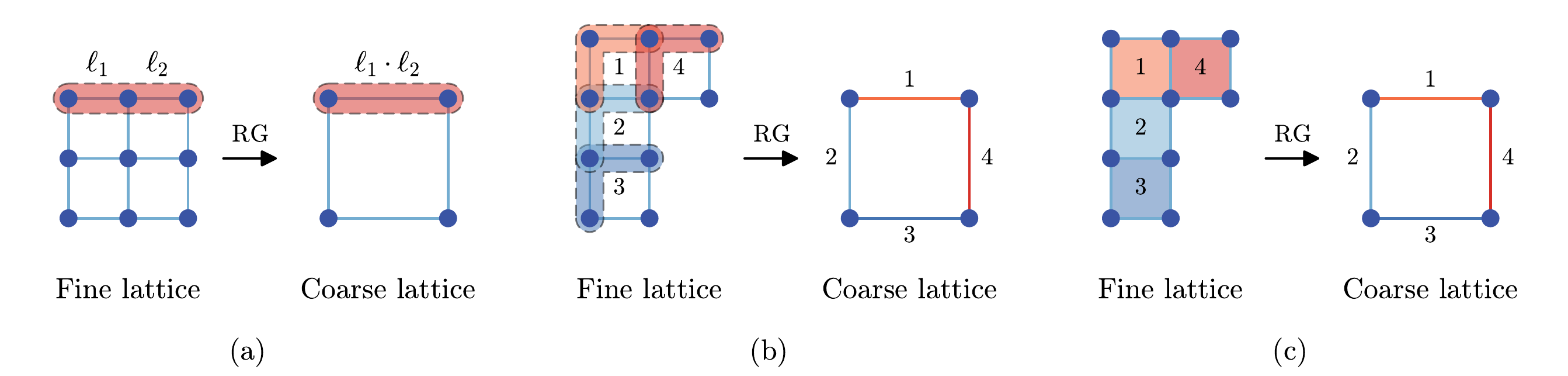}\label{fig:schematic_links}}
\caption{%
    \textbf{RG schemes.}
    (a) The standard scheme used sends products of gauge links to another gauge link in the same direction, preserving link covariance.
    (b)In an alternating scheme we send the product of top and left links of the fine lattice to the top and left links of the coarse lattice sequentially. 
    (c)Another alternating rule sends plaquettes sequentially to the top and left links; under periodic boundary conditions, this two-step mapping covers every link uniformly.
  }
\end{figure}

\noindent In this appendix, we show that the real‐space coarse‐graining prescription shown in Fig.~\hyperref[fig:schematic_links]{3a} applied for this model—as well as for the Schwinger and mixed models preserves gauge invariance. We work on a square lattice with gauge group $G$ and link variables
$U_{x,\mu}\in G$ on directed links $(x,\mu)$, $\mu\in\{\hat x,\hat y\}$.
A local gauge transformation $\{g_x\in G\}$ acts as
\begin{equation}
U_{x,\mu}\;\longmapsto\; g_x\,U_{x,\mu}\,g_{x+\hat\mu}^{-1}.
\label{eq:gauge-transform}
\end{equation}
\noindent \emph{Blocking map.}
Define coarse sites on the even sublattice $X=(2i,2j)$. The coarse links are straight
length–2 products of fine links in the same direction:
\begin{equation}
U'_{X,\hat x}\;:=\;U_{X,\hat x}\,U_{X+\hat x,\hat x},
\qquad
U'_{X,\hat y}\;:=\;U_{X,\hat y}\,U_{X+\hat y,\hat y}.\\
\label{eq:blocking-def}
\end{equation}

\noindent \emph{Claim 1 (Gauge covariance of coarse links).}
Under \eqref{eq:gauge-transform},
\begin{align}
U'_{X,\hat x}&\longmapsto
\bigl(g_X U_{X,\hat x} g_{X+\hat x}^{-1}\bigr)
\bigl(g_{X+\hat x} U_{X+\hat x,\hat x} g_{X+2\hat x}^{-1}\bigr)
= g_X\,U'_{X,\hat x}\,g_{X+2\hat x}^{-1},\nonumber\\
U'_{X,\hat y}&\longmapsto g_X\,U'_{X,\hat y}\,g_{X+2\hat y}^{-1}.
\label{eq:coarse-covariance}
\end{align}
Thus with coarse gauge factors $G_X:=g_X$ we have
$U'_{X,\mu'}\mapsto G_X\,U'_{X,\mu'}\,G_{X+\hat\mu'}^{-1}$ (where $\hat{\mu'} = 2\hat{\mu}$ is the unit vector of the coarse lattice in direction $\mu$),
the coarse links preserve gauge-covariance in the blocked lattice.\\

\noindent \emph{Claim 2 (Gauge invariance of coarse plaquettes and Wilson loops).}
Define the coarse $1\times1$ plaquette at $X$ by
\begin{equation}
P'_X\;:=\;U'_{X,\hat x'}\,U'_{X+\hat x',\hat y'}\,
\bigl(U'_{X+\hat y',\hat x'}\bigr)^{-1}\,\bigl(U'_{X,\hat y'}\bigr)^{-1}.
\label{eq:coarse-plaquette}
\end{equation}
From \eqref{eq:coarse-covariance} it follows that
$P'_X\mapsto G_X\,P'_X\,G_X^{-1}$,
so $P'_X$ is gauge invariant for Abelian $G$ (e.g.\ $\mathbb Z_2$, $U(1)$),
and $\mathrm{Tr}\,P'_X$ is gauge invariant for non–Abelian $G$.
Any coarse Wilson loop (product of coarse links around a closed contour)
is likewise gauge invariant.\\

\noindent \emph{Local relation to fine plaquettes.}
The fine $1\times1$ plaquette based at $x$ is
\begin{equation}
P_x:=U_{x,\hat x}\,U_{x+\hat x,\hat y}\,U^{-1}_{x+\hat y,\hat x}\,U^{-1}_{x,\hat y}.
\label{eq:fine-plaquette}
\end{equation}
Inserting \eqref{eq:blocking-def} into \eqref{eq:coarse-plaquette} shows that $P'_X$
is the ordered product of the four fine plaquettes in the $2\times2$ block with
corner $X$, so that $\mathrm{Tr}\,P'_X=\mathrm{Tr}\!\bigl(P_X P_{X+\hat x} P_{X+\hat y} P_{X+\hat x+\hat y}\bigr)$ is gauge invariant.
Hence the blocking is local and preserves gauge–invariant observables. This construction generalizes immediately to arbitrary compact groups in $d$ dimensions by multiplying consecutive links along  the $d$ coordinate directions.\\

\noindent For this model, we also applied the blocking procedures illustrated in Figs.~\hyperref[fig:schematic_links]{3b} and \hyperref[fig:schematic_links]{3c}. We find that in every case, the effective coupling $K$ is driven monotonically to $K=0$, the trivial fixed point of the theory. Since plaquettes of the fine lattice are mapped to links of the coarse lattice, these schemes do not preserve the covariance at the link level. However, they still preserve the local gauge–invariant content of a $2\times 2$ block (small Wilson loops) for Abelian Theories. Thus they produce renormalized coupling $K'(K)$ that is in agreement from theoretical decimation schemes. Each of these real–space RG constructions admits a straightforward extension to higher dimensions.\\

\section{Score Function for the Schwinger Model}
\label{sec:sf_sm}
\noindent The Euclidean action of the Schwinger Model is
\begin{align*}
S = S_{\text{gauge}} + \psi^\dagger M \psi,
\qquad
S_{\text{gauge}} = -\sum_{\text{plaq}} \beta\bigl(1-\cos\theta_{\text{plaq}}\bigr),
\end{align*}
where the operator $M$ acts on the fermion fields $\psi(x)$
\begin{align*}
M\psi(x) = (m+2)\psi(x)
-\frac{1}{2}\sum_{\mu=1}^{2} \Big[ U_\mu(x)(1-\gamma_\mu)\psi(x+\hat\mu)
+ U^\dagger_\mu(x-\hat\mu)(1+\gamma_\mu)\psi(x-\hat\mu)\Big],
\end{align*}
and
\begin{align*}
\theta_{\text{plaq}}(i)=\theta_x(i)+\theta_t(i_{\text{right}})-\theta_x(i_{\text{up}})-\theta_t(i).
\end{align*}

\subsection*{First Derivatives}
\noindent Gauge part:
\begin{align*}
S'_{\theta_x} = \frac{\partial S}{\partial \theta_x(i)} &=
-\beta\left[\sin\theta_{\text{plaq}}(i_{\text{down}})-\sin\theta_{\text{plaq}}(i)\right],\\
S'_{\theta_t} =\frac{\partial S}{\partial \theta_t(i)} &=
-\beta\left[\sin\theta_{\text{plaq}}(i_{\text{left}})-\sin\theta_{\text{plaq}}(i)\right].
\end{align*}
Fermion part via pseudofermions $\phi$:
\begin{align*}
S_{\text{pf}} = \phi^\dagger (MM^\dagger)^{-1}\phi,
\qquad
\Psi \equiv M^{-1}\phi\ \Rightarrow\ S_f=\Psi^\dagger\Psi.
\end{align*}
Varying $S_f$ with respect to link angles,
\begin{align*}
\delta S_f = -2\,\Re\!\left(\Psi^\dagger M^{-1}\delta M\,\Psi\right),
\qquad
S_f' = -2\,\Re\!\left(\Psi^\dagger M^{-1} M' \Psi\right).
\end{align*}
where $M'$ is the derivative of $M$. The Dirac operator elements:
\begin{align*}
M_{n\alpha,m\beta} &= (m+2)\delta_{nm}\delta_{\alpha\beta}
-\frac{1}{2}\sum_{\mu=1}^2\Big[(1-\gamma_\mu)_{\alpha\beta}U_{n,\mu}\delta_{n,m-\hat\mu}
+(1+\gamma_\mu)_{\alpha\beta}U^\dagger_{m,\mu}\delta_{n,m+\hat\mu}\Big],\\
M^{(1,\theta_x)}_{n\alpha,m\beta} &= -\frac{i}{2}\Big[(1-\gamma_1)_{\alpha\beta}U_{n,1}\delta_{n,m-\hat1}
-(1+\gamma_1)_{\alpha\beta}U^\dagger_{m,1}\delta_{n,m+\hat1}\Big],
\end{align*}
make the indices explicit. The matrix $M^{(1,\theta_x)}$ denotes the first derivative of $M$ w.r.t $\theta_x$ and the extension for $\theta_t$ is straightfoward. 
\\
\subsection*{Second Derivatives}
\noindent Gauge part:
\begin{align*}
S''_{\theta_x} = \frac{\partial^2 S}{\partial \theta_x(i)^2} &=
-\beta\left[\cos\theta_{\text{plaq}}(i_{\text{down}})+\cos\theta_{\text{plaq}}(i)\right],\\
S''_{\theta_t} =\frac{\partial^2 S}{\partial \theta_t(i)^2} &=
-\beta\left[\cos\theta_{\text{plaq}}(i_{\text{left}})+\cos\theta_{\text{plaq}}(i)\right].
\end{align*}
\noindent Starting from
\begin{align*}
\delta^2 S_f = -2\,\Re\big(\delta(\Psi^\dagger M^{-1} M' \Psi)\big),
\end{align*}
and using $\delta\Psi = -M^{-1}\delta M\,\Psi$,
\begin{align*}
S_f'' =
\Psi^\dagger\!\left(
- M'^\dagger (M^\dagger)^{-1}M^{-1}M'
- 2\,M^{-1}M' M^{-1} M'
+ M^{-1}M''
\right)\!\Psi.
\end{align*}
Here $M''\equiv\partial^2 M/\partial \theta^2$, e.g.
\begin{align*}
M^{(2,\theta_x)}_{n\alpha,m\beta}
=\frac{1}{2}\Big[(1-\gamma_1)_{\alpha\beta}U_{n,1}\delta_{n,m-\hat1}
+ (1+\gamma_1)_{\alpha\beta}U^\dagger_{m,1}\delta_{n,m+\hat1}\Big],
\end{align*}
and analogously for $\theta_t$. Thus the score function for the gauge part of the action is
\begin{align*}
    \text{Score}_{\theta} = \sum_{T=1}^{N_{\text{samp}
    }}\sum_{\theta_x, \theta_t}-S_f'' + \frac{1}{2}(S_f')^2 
\end{align*}
where the derivative denoted by prime is with respect to $\theta_x$ and $\theta_t$.
\subsection*{Score Function derivatives w.r.t. fermions}
\noindent For real variables $\xi_i$ sampled from $S$, the score estimator is
\begin{align*}
\text{Score} = \sum_{T=1}^{N_{\text{samp}}} \sum_i \left[
-\frac{\partial^2 S}{\partial \xi_i^2}
+\frac{1}{2}\left(\frac{\partial S}{\partial \xi_i}\right)^2
\right].
\end{align*}
For complex variables $z=x+iy$, $\bar z=x-iy$,
\begin{align*}
\frac{\partial S}{\partial z} &= \frac{1}{2}\left(\frac{\partial S}{\partial x}-i\frac{\partial S}{\partial y}\right),\quad
\frac{\partial S}{\partial \bar z} = \frac{1}{2}\left(\frac{\partial S}{\partial x}+i\frac{\partial S}{\partial y}\right),\\
\frac{\partial^2 S}{\partial z\,\partial \bar z}
&= \frac{1}{4}\left(\frac{\partial^2 S}{\partial x^2}+\frac{\partial^2 S}{\partial y^2}\right).
\end{align*}
If $\partial S/\partial \bar z = \overline{\partial S/\partial z}$ (which is true for $S_f$), then the score loss is
\begin{align*}
\text{Score} = \sum_{T=1}^{N_{\text{samp}}}-4\frac{\partial^2 S}{\partial z\,\partial \bar z}
+ 2\left\lVert \frac{\partial S}{\partial z}\right\rVert^2.
\end{align*}
For the pseudofermion part
\begin{align*}
S_f = \sum_{ij}\phi_i^* (MM^\dagger)^{-1}_{ij}\phi_j
= \phi^\dagger (MM^\dagger)^{-1}\phi,
\end{align*}
we obtain
\begin{align*}
\frac{\partial S_f}{\partial \phi_i^*}
= \sum_j (MM^\dagger)^{-1}_{ij}\phi_j
= \gamma_5 Q^{-1}\Psi,
\qquad Q\equiv \gamma_5 M=Q^\dagger,
\end{align*}
so that
\begin{align*}
\sum_i \left\lVert\frac{\partial S_f}{\partial \phi^*_i}\right\rVert^2
= \left\lVert \gamma_5 Q^{-1}\Psi\right\rVert^2,
\qquad
\sum_i \frac{\partial^2 S_f}{\partial \phi_i^* \partial \phi_i}
= \mathrm{Tr}\,(MM^\dagger)^{-1}
= \mathrm{Tr}\,Q^{-2}.
\end{align*}
Thus the score loss for pseudofermions is
\begin{align*}
\text{Score}_f = \sum_{T=1}^{N_{\text{samp}}}\left(-4\,\mathrm{Tr}\,Q^{-2}
+ 2\left\lVert \gamma_5 Q^{-1}\Psi\right\rVert^2\ \right)
\end{align*}
The final score loss will be
\begin{align*}
    \text{Final Score} &= \text{Score}_{\theta} + \text{Score}_{f}\\
    &= \sum_{T=1}^{N_{\text{samp}
    }}\sum_{\theta_x, \theta_t}\left(-S_\theta'' + \frac{1}{2}(S_\theta')^2 \right) + \left(-S_f'' + \frac{1}{2}(S_f')^2 \right)+ \left(-4\,\mathrm{Tr}\,Q^{-2}
+ 2\left\lVert \gamma_5 Q^{-1}\Psi\right\rVert^2\ \right)\;.
\end{align*}
\end{document}